\documentclass[sigconf]{acmart}
\usepackage{colortbl}
\usepackage{subcaption}

\AtBeginDocument{%
  }

\copyrightyear{2026}
\acmYear{2026}
\setcopyright{cc}
\setcctype{by}
\acmConference[ICPC '26]{34th IEEE/ACM International Conference on Program Comprehension}{April 12--13, 2026}{Rio de Janeiro, Brazil}
\acmBooktitle{34th IEEE/ACM International Conference on Program Comprehension (ICPC '26), April 12--13, 2026, Rio de Janeiro, Brazil}
\acmPrice{}
\acmDOI{10.1145/3794763.3794799}
\acmISBN{979-8-4007-2482-4/2026/04}

\begin{document}

\title[EyeLayer]{EyeLayer: Integrating Human Attention Patterns into LLM-Based Code Summarization}

\author{Jiahao Zhang}
\email{jiahao.zhang@vanderbilt.edu}
\orcid{0009-0008-8379-6871}
\affiliation{%
  \institution{Vanderbilt University}
  \city{Nashville}
  \state{Tennessee}
  \country{USA}}

\author{Yifan Zhang}
\email{yifan.zhang.2@vanderbilt.edu}
\affiliation{%
  \institution{Vanderbilt University}
  \city{Nashville}
  \state{Tennessee}
  \country{USA}}

\author{Kevin Leach}
\email{kevin.leach@vanderbilt.edu}
\affiliation{%
  \institution{Vanderbilt University}
  \city{Nashville}
  \state{Tennessee}
  \country{USA}}

\author{Yu Huang}
\email{yu.huang@vanderbilt.edu}
\affiliation{%
  \institution{Vanderbilt University}
  \city{Nashville}
  \state{Tennessee}
  \country{USA}}

\renewcommand{\shortauthors}{Zhang et al.}
\begin{abstract}
Code summarization is the task of generating natural language descriptions of source code, which is critical for software comprehension and maintenance. While large language models (LLMs) have achieved remarkable progress on this task, an open question remains: can human expertise in code understanding further guide and enhance these models? We propose EyeLayer, a lightweight attention-augmentation module that incorporates human eye-gaze patterns, as a proxy of human expertise, into LLM-based code summarization. EyeLayer models human attention during code reading via a Multimodal Gaussian Mixture, redistributing token embeddings based on learned parameters $(\mu_i, \sigma_i^2)$ that capture where and how intensively developers focus. This design enables learning generalizable attention priors from eye-tracking data and incorporating them into LLMs seamlessly, without disturbing existing representations. We evaluate EyeLayer across diverse model families (i.e., LLaMA-3.2, Qwen3, and CodeBERT) covering different scales and architectures. EyeLayer consistently outperforms strong fine-tuning baselines across standard metrics, achieving gains of up to 13.17\% on BLEU-4. These results demonstrate that human gaze patterns encode complementary attention signals that enhance the semantic focus of LLMs and transfer effectively across diverse models for code summarization.
\end{abstract}

\ccsdesc[500]{Software and its engineering~Software creation and management}
\ccsdesc[500]{Computing methodologies~Artificial intelligence}


\keywords{Code Summarization, Human Factors in Software Engineering, Human-centered AI for Software Engineering}


\maketitle
\section{Introduction}

Software documentation is an essential bridge between code implementation and developer understanding, with code summarization facilitating efficient program comprehension~\cite{ahmad2020summarization,sun2024source}. As modern software systems become increasingly complex, quickly grasping code functionality through concise summaries is critical for maintenance and evolution. Consequently, automatically generating high-quality summaries has become a central challenge in software engineering~\cite{karas2024tale,rodegheroImprovingAutomatedSource2014b}.

Recent advances in large language models (LLMs) have demonstrated remarkable capabilities in code-related tasks, particularly in code summarization~\cite{sun2024source,hou2024largelanguagemodelssoftware,fan_llm4se}. While these models have achieved good performance by learning from vast corpora of code–summary pairs, there remains a gap in generating human-aligned summaries that capture the information humans actually focus on during code comprehension\cite{bansalModelingHumanAttention2023b,alakmehPredictingCodeComprehension2024a}. Meanwhile, when developers comprehend code to formulate summaries, their attention patterns reveal how they selectively allocate focus across different parts of the code~\cite{karas2024tale,sharafiPracticalGuideConducting2020b}. In previous software engineering research, eye-tracking studies have been widely used to extract developers' attention patterns which is a promising proxy for their cognitions during programming activities~\cite{sharifEyeTrackingStudy2010a,sharafiEyetrackingMetricsSoftware2015,sharafiSystematicLiteratureReview2015}. This motivates a key question: \textbf{can incorporating human attention signals further enhance LLM-based code summarization?}

The most recent research has attempted to guide AI model development leveraging developers' attention patterns and demonstrated promising benefits of such guidance. EyeTrans~\cite{Eyetrans} for the first time integrated eye-gaze signals into a single Transformer block for code summarization, achieving up to 6.39\% improvement. However, it remains unknown whether human attention can actually enhance modern LLMs, which differ substantially in scale, architecture, and optimization dynamics. This uncertainty limits their potential impact on real-world applications.

To bridge the gap between human and LLM attention mechanisms, we propose \textbf{EyeLayer}, a lightweight architectural module that integrates human eye-gaze data into LLM-based code summarization. Our approach is grounded in a key insight: during code comprehension, programmers naturally focus their attention unevenly across the code, concentrating intensively on semantically critical regions while peripherally attending to contextual elements. EyeLayer models this distributional attention as a transferable prior, learned from a curated eye-tracking corpus of 27 professional developers~\cite{Eyetrans}, which captures how human gaze behavior reflects semantic importance during real code comprehension. It employs a \textbf{Multimodal Gaussian Mixture} to redistribute each code embedding based on learned parameters $(\mu_i, \sigma_i^2)$, which encode both the intensity and spread of human attention. Integrated into the supervised fine-tuning process, EyeLayer leverages these human-derived priors to improve how pretrained models allocate focus across code tokens without altering the original model architecture. Despite being trained on a small but cognitively grounded dataset, EyeLayer generalizes effectively to large-scale LLMs, showing that even limited human attention data can yield measurable improvements.

Functionally, EyeLayer serves as a recommendation system for code embedding redistribution: for each code embedding, it predicts a small set of Gaussian modes that recommend how its representation should be redistributed. This mechanism allows the model to compose fine-grained and global focus patterns, analogous to personalized recommendation in representation space. By decoupling gaze-informed redistribution from the model’s intrinsic attention weights, EyeLayer learns generalizable attention priors from sparse eye-tracking data and transfers them to unseen code. Incorporated within LLMs, it preserves pretrained representations while infusing human-like focus behavior directly into the attention redistribution process.

We evaluate EyeLayer across five models spanning different scales and architectures: CodeBERT (125M)~\cite{fengCodeBERTPretrainedModel2020}, LLaMA-3.2-1B/3B-Instruct~\cite{grattafioriLlama3Herd2024a}, and Qwen3-1.7B/4B-base~\cite{yangQwen3TechnicalReport2025}. All EyeLayer-augmented models are compared against strong supervised fine-tuned baselines trained on identical code summarization data (CodeXGLUE~\cite{codexglue,husain2019codesearchnet}) but without eye-tracking integration, isolating the contribution of human attention signals. Evaluation uses four widely-adopted metrics capturing lexical overlap (BLEU\cite{papineniBleuMethodAutomatic2002a}, ROUGE-L\cite{linROUGEPackageAutomatic2004}, METEOR\cite{banerjeeMETEORAutomaticMetric2005}) and semantic similarity (BERTScore\cite{zhangBERTScoreEvaluatingText2020}). Across all five models, EyeLayer achieves consistent gains over fine-tuning baselines, with improvements up to 13.17\% on BLEU-4, confirming that human attention signals enhance LLM performance across architectures.

This paper makes the following contributions:
\begin{itemize}
    \item We propose a framework for integrating human cognitive priors into large language models for code summarization. Using eye-tracking data as transferable probabilistic priors, our approach establishes a bridge between human attention behavior and LLM-level attention formation.
    \item We design the \textit{Multimodal Gaussian EyeLayer}, a lightweight, recommendation-like module that redistributes code embeddings through learnable Gaussian mixtures. This mechanism decouples gaze-informed redistribution from intrinsic attention weights, enabling scalable integration of sparse human signals into billion-parameter LLMs.
    \item We conduct a systematic evaluation across five LLMs spanning both encoder-only and decoder-only architectures, demonstrating consistent improvements on the CodeXGLUE benchmark and strong transferability of learned attention priors to unseen code.
    \item To facilitate reproducibility and foster future research, we release our implementation scripts and datasets at \href{https://zenodo.org/records/17452570?token=eyJhbGciOiJIUzUxMiJ9.eyJpZCI6ImZhZjA1YjA5LWEzZTItNDA2My04NGE0LTJiMjQ5Mzg1OGVlMiIsImRhdGEiOnt9LCJyYW5kb20iOiJlNmQzOTZlMWE2ZmJkNmFlMjRjMmU1NTliYWZlMjc5NSJ9.35itoTvmSoTdm-NW2Eibx6E7OO50lpoFzZxBcDS2VJu3aKilXL60f4JkjVNoz7kMkTIfvSM8AD-yo9qYTrN5ng}{\textbf{URL}}.
\end{itemize}


In the rest of this paper, Section~\ref{sec:background} presents the background of eye-tracking in program comprehension and probabilistic attention modeling. 
Section~\ref{sec:methodology} introduces the design and implementation details of the proposed \textit{EyeLayer} architecture.
Section~\ref{sec:experiment_setup} details the experimental setup.
Section~\ref{sec:results} analyze the results. 
Section~\ref{sec:threats} discusses potential threats to validity. 
Section~\ref{sec:discussion} provides a broader discussion of findings and implications. 
Section~\ref{sec:relatedwork} reviews related work. 
Finally, Section~\ref{sec:conclusion} concludes the paper and outlines directions for future research.
\section{Background}\label{sec:background}
Human gaze behavior offers empirical insight into how developers comprehend code, while probabilistic attention provides a principled way to model such focus computationally. This section reviews key findings from eye-tracking studies and links them to Gaussian-based attention formulations that inspire our EyeLayer design.

\subsection{Eye-tracking for Program Comprehension}

Eye-tracking has become a rigorous method for examining cognitive processes in software engineering research, particularly in understanding how developers read and comprehend source code~\cite{grabingerEyeTrackingSoftware2024,sharafiPracticalGuideConducting2020b}. By capturing gaze behavior, eye-tracking enables the quantitative analysis of attention allocation and processing effort with high temporal precision. In software engineering, this relationship is particularly relevant because program comprehension, like natural language reading, involves the incremental interpretation of complex visual and semantic structures~\cite{sharafiSystematicLiteratureReview2015}. Fixation-based metrics provide a means to infer where and when developers engage in information processing, distinguishing meaningful cognitive activity from mere visual transitions represented by saccades~\cite{sharafiEyetrackingMetricsSoftware2015}.

The theoretical basis for interpreting gaze data originates from cognitive psychology, most notably the work of Just and Carpenter~\cite{justTheoryReadingEye}. Their eye–mind assumption states that the duration of a fixation, the period of relative ocular stability directly reflects the time required for cognitive processing. This principle established fixations as a reliable indicator of comprehension effort in reading, linking visual attention to linguistic and semantic processing. Empirical evidence shows that fixations occupy the vast majority of viewing time during code reading, emphasizing their role as the fundamental unit of analysis for understanding comprehension behavior~\cite{sharafiPracticalGuideConducting2020b,sharafiSystematicLiteratureReview2015}. Overall, fixation analysis offers a direct and interpretable connection between observable gaze patterns and the underlying cognitive mechanisms of program understanding, making eye-tracking a valuable empirical approach for investigating how developers read, reason about, and make decisions based on source code.

\subsection{Probabilistic Attention and Cognitive Priors}

Transformer attention can be framed probabilistically, with weights parameterized as continuous distributions over positions. Gaussian parameterizations offer a simple and interpretable form: a mean for focus location and a variance for spread. Representative studies show concrete uses of such priors. Chorowski et al.\ introduced Gaussian-shaped attention for sequence-to-sequence alignment in speech recognition~\cite{chorowskiAttentionBasedModelsSpeech2015}. Cordonnier et al.\ analyzed self-attention and showed that learned patterns relate closely to Gaussian-like kernels over relative positions~\cite{cordonnierRelationshipSelfAttentionConvolutional2019}. You et al.\ further reported that hard-coded Gaussian windows can match the performance of fully learned attention in machine translation, indicating that Gaussian structure can serve as an effective bias~\cite{youHardCodedGaussianAttention2020}. To allow multiple foci, Graves modeled attention as a mixture of Gaussians in recurrent architectures, capturing multi-modal alignments with learnable centers and spreads~\cite{gravesGeneratingSequencesRecurrent2014}.

This probabilistic view aligns with findings from eye-tracking. Studies in software engineering report localized and selective fixations during code reading~\cite{sharifEyeTrackingStudy2010a, busjahnEyeMovementsCode2015b}. Such fixation maps are commonly summarized as peaked distributions over spatial locations. Neural models inspired by selective vision, such as DRAW, use parameterized Gaussian filters to realize differentiable focus regions~\cite{gregorDRAWRecurrentNeural2015}. These results motivate representing model attention with Gaussian or mixture forms when human-like focus is desirable.

Guided by this evidence, our EyeLayer treats attention as a learnable mixture with sparse mode selection. The formulation provides an interpretable parameter space (centers, spreads, and weights) consistent with probabilistic attention and with observed fixation patterns in code comprehension. This connects a statistical prior on attention with cognitively grounded signals in a single mechanism.

\section{Methodology}\label{sec:methodology}

\begin{figure*}[t]
    \centering
    \includegraphics[width=\textwidth]{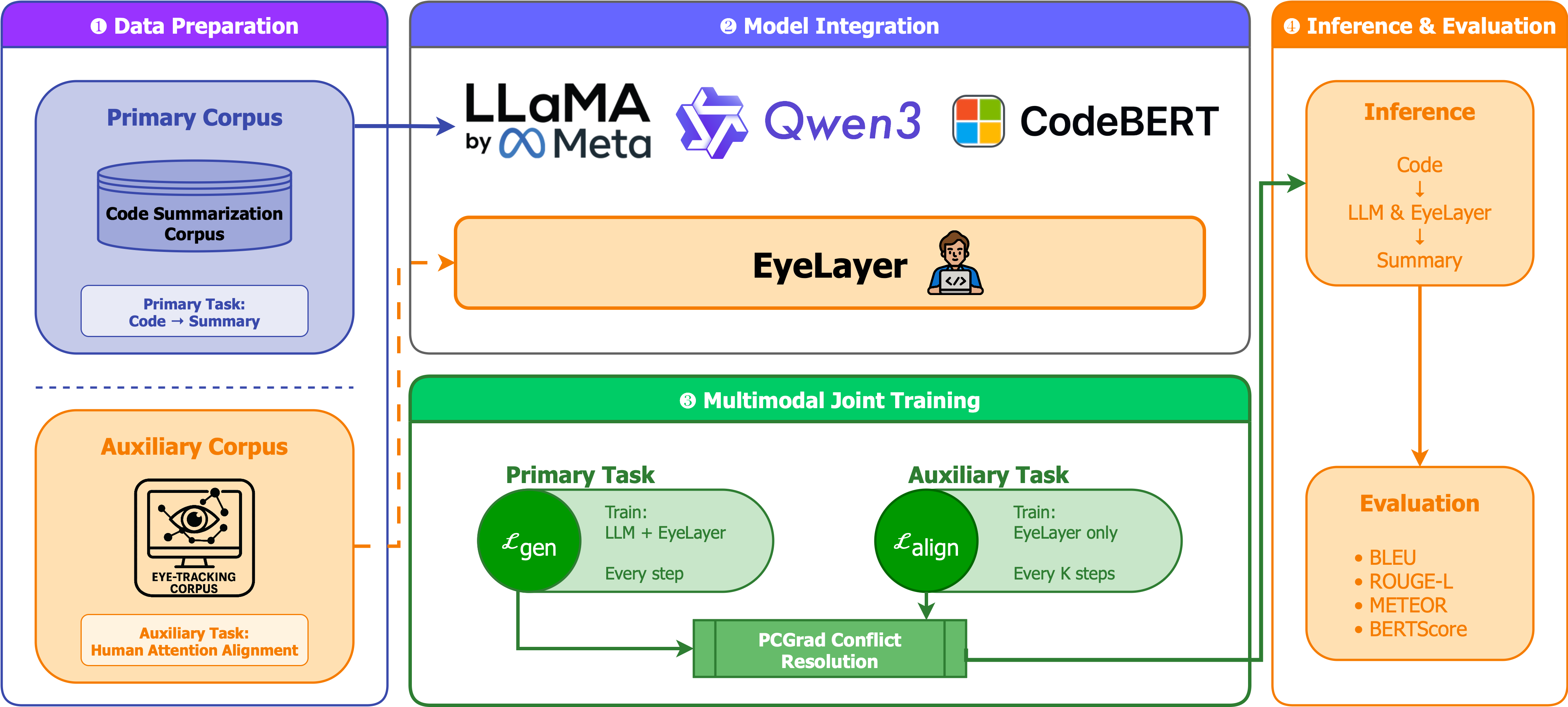}
    \caption{Overview of our joint training pipeline.}
    \Description{The figure shows the joint training pipeline that combines a code summarization task with an auxiliary eye-tracking alignment task through the shared EyeLayer.}
    \label{fig:pipeline}
\end{figure*}

Our training employs a carefully designed dual-dataset strategy that separates the primary code summarization task from the auxiliary eye-tracking alignment task, enabling learning from both large-scale code-summary pairs and sparse but cognitively grounded attention signals. Figure~\ref{fig:pipeline} provides an overview of our complete approach.

\subsection{Datasets and Preprocessing}


\textbf{Code Summarization Corpus.} We use the Java subset of the CodeXGLUE benchmark~\cite{codexglue}, a widely-adopted dataset containing Java methods paired with their corresponding docstring summaries extracted from open-source repositories. The dataset provides diverse code patterns spanning different programming idioms, complexity levels, and documentation styles, enabling robust learning of the code-to-summary mapping across varied contexts.


\textbf{Eye-Tracking Corpus.}  
We derive the auxiliary alignment supervision from the EyeTrans corpus~\cite{Eyetrans}, which records the gaze behaviors of developers during controlled code comprehension tasks. Each sample links a Java method to its Abstract Syntax Tree (AST) and corresponding fixations that capture how programmers allocate attention across syntactic and semantic regions. A \textit{fixation} is defined as a spatially stable gaze lasting approximately 100--300\,ms~\cite{sharafiPracticalGuideConducting2020b}, during which most visual information processing occurs~\cite{justTheoryReadingEye}. Each fixation is localized on screen coordinates and mapped to its corresponding AST node, producing discrete yet cognitively grounded attention signals. These node-level fixation counts are then aligned to model-level subtoken representations through our three-stage matching pipeline described below, providing precise human-derived supervision for multimodal alignment in the EyeLayer.

To effectively integrate these fixation-based signals into the model, we must reconcile the representational gap between the human gaze space and the model input space. The eye-tracking corpus encodes attention in the \textit{AST node space}, identifying which syntactic constructs programmers focus on, whereas the multimodal EyeLayer operates in the \textit{subtoken space}, defined by byte-pair encoded tokens from the model tokenizer. This mismatch is non-trivial: (1) a single AST node like \texttt{BFSdistance} may split into multiple subtokens [\texttt{BFS}, \texttt{distance}], (2) tokenization varies based on surrounding context and instruction templates, and (3) abstract AST nodes have no direct token correspondence. We address this through a three-stage alignment pipeline: first, we traverse the AST to extract concrete code elements; second, we apply context-aware tokenization matching the model's instruction format; finally, we use multi-strategy matching—exact matching for simple cases, consecutive aggregation for split tokens, and character offset estimation for complex constructs—to map AST nodes to subtoken indices. This pipeline achieves >98\% mapping accuracy and enables us to transfer sparse node-level fixation counts to the dense subtoken representations required for attention supervision.

\textbf{Independent Data Sources.} To ensure clear supervision boundaries, the two datasets are kept entirely independent. The code summarization corpus drives the primary generation objective, while the eye-tracking corpus contributes auxiliary alignment supervision. They contain disjoint code samples, eliminating data leakage and ensuring that observed improvements stem from the integration of human cognitive priors rather than exposure to additional labeled summaries.

\subsection{Multimodal Gaussian EyeLayer}

\begin{figure}[htbp]
    \centering
    \includegraphics[width=0.4\textwidth]{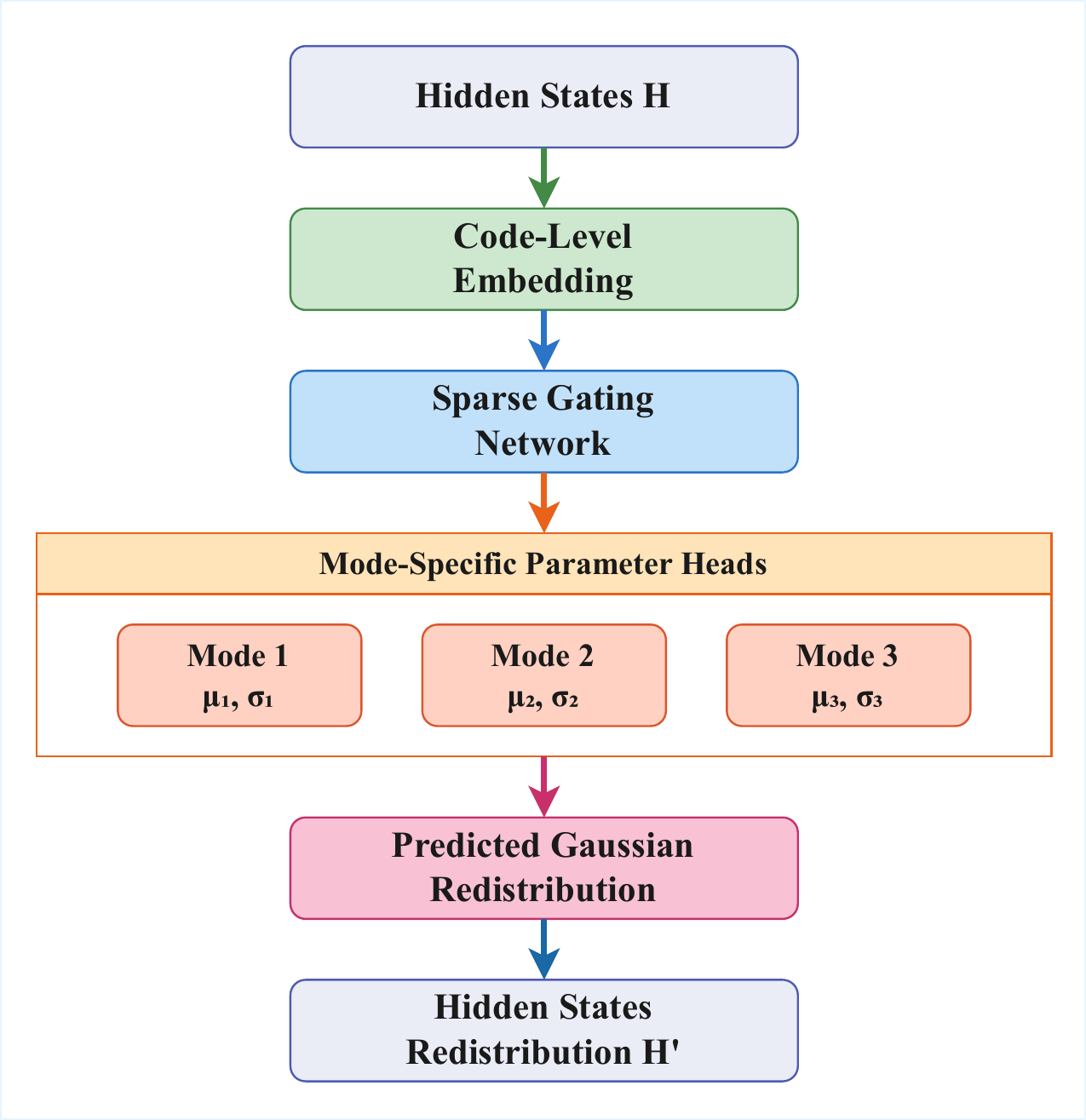}
    \caption{The Multimodal Gaussian EyeLayer architecture.}
    \Description{The figure illustrates the Multimodal Gaussian EyeLayer, including code embedding, sparse gating, Gaussian parameter prediction, and mixture-based attention generation.}
    \label{fig:architecture}
\end{figure}


Our approach builds on the key insight that, during code comprehension, programmers allocate attention unevenly across the code: they concentrate intensively on semantically critical regions while attending peripherally to contextual elements.  
This uneven distribution can be viewed as a composition of several focus patterns, each representing a localized concentration of attention over the token sequence.  
To model this behavior, the \textbf{Multimodal Gaussian EyeLayer} represents attention as a mixture of Gaussian components.  
Each component defines a focus region characterized by a center $\mu_k$ (semantic locus) and spread $\sigma_k$ (contextual extent), while a sparse gating network determines how many such regions are needed for each code snippet.  
This formulation captures both concentrated and distributed focus within a unified probabilistic framework, allowing the model to adaptively modulate attention according to code structure and semantics.

The EyeLayer integrates into pretrained decoder-only transformers (e.g., LLaMA, Qwen) through hook-based injection at an intermediate layer.  
During forward propagation, it intercepts hidden states $\mathbf{H}$, applies the EyeLayer transformation, and returns updated representations $\mathbf{H}'$ to subsequent layers.  
This hook-based design preserves the causal structure of the base model while enriching its intermediate representations with human-aligned attention priors, as illustrated in Figure~\ref{fig:architecture}.







\subsubsection{Code-Level Embedding}

Before predicting Gaussian parameters, the model first summarizes the overall semantic context of the input sequence.  
For hidden states $\mathbf{H} \in \mathbb{R}^{B \times L \times d}$ from an intermediate transformer layer, we apply an attention mask $\mathbf{M}_{\text{attn}}$ and a special-token mask $\mathbf{M}_{\text{special}}$ to form $\mathbf{M} = \mathbf{M}_{\text{attn}} \odot (1 - \mathbf{M}_{\text{special}})$.  
When positional information is available, a decay factor $D_{p_i} = \gamma^{p_i}$ ($\gamma=0.95$) down-weights distant tokens.  
The code-level embedding is computed as:
\begin{equation}
\mathbf{e} = \frac{\sum_{i=1}^{L} M_i D_{p_i} \mathbf{H}_i}{\sum_{i=1}^{L} M_i + \epsilon},
\end{equation}
where $\epsilon$ is a small constant to avoid division by zero.  
The resulting vector $\mathbf{e} \in \mathbb{R}^{d}$ provides a compact semantic summary for mode prediction.




\subsubsection{Sparse Gating Mechanism}

To decide how many Gaussian components should be activated for each code sequence, the EyeLayer uses a lightweight gating network that maps the code embedding $\mathbf{e}$ to a normalized weight vector $\mathbf{w} \in \mathbb{R}^K$:
\begin{equation}
\mathbf{w} = \operatorname{softmax}\!\left(\mathbf{W}_2\,\phi(\mathbf{W}_1\mathbf{e}+\mathbf{b}_1)+\mathbf{b}_2\right),
\end{equation}
where $\phi(\cdot)$ is a non-linear activation, and $\mathbf{W}_1,\mathbf{W}_2,\mathbf{b}_1,\mathbf{b}_2$ are learnable projection and bias parameters.  
The softmax normalization ensures $\sum_k w^{(k)}=1$, yielding interpretable mode activations that indicate the relative contribution of each Gaussian component.  
This gating mechanism encourages sparse activation: simple functions tend to concentrate weight on a single mode, whereas more complex code distributes attention across multiple regions.  
Such adaptive allocation allows the model to adjust its focus continuously without introducing discrete decisions or additional supervision.



\subsubsection{Mode-Specific Parameterization}

Each active mode predicts its Gaussian parameters based on shared semantic features extracted from the same code embedding $\mathbf{e}$.  
The shared representation is computed as:
\begin{equation}
\mathbf{h}_{\text{shared}} = \text{Dropout}(\text{LayerNorm}(\text{GELU}(\mathbf{W}_h \mathbf{e} + \mathbf{b}_h))),
\end{equation}
where $\mathbf{W}_h$ and $\mathbf{b}_h$ are learnable projection parameters.  
Each mode then applies lightweight linear heads:
\begin{align}
\tilde{\mu}_k &= \mathbf{w}_{\mu}^{(k)} \mathbf{h}_{\text{shared}} + b_{\mu}^{(k)}, \\
\tilde{\sigma}_k &= \mathbf{w}_{\sigma}^{(k)} \mathbf{h}_{\text{shared}} + b_{\sigma}^{(k)},
\end{align}
where $\tilde{\mu}_k$ and $\tilde{\sigma}_k$ are raw predictions for the center and spread of the $k$-th Gaussian component.  
Predictions are constrained to $\mu_k \in [0, L{-}1]$ and $\sigma_k \in [\sigma_{\min}, L/2]$ to ensure valid ranges.  
Centroid biases are initialized to cover early, middle, and late regions of the sequence to promote spatial diversity during early training.






\subsubsection{Gaussian Mixture Construction}\label{sec:gaussian_mix}

The final attention distribution is formed as a weighted mixture of $K{=}3$ Gaussian components:
\begin{equation}
P(i) = \sum_{k=1}^{K} w^{(k)}
\frac{\exp\!\left(-\frac{(i - \mu_k)^2}{2\sigma_k^2}\right)}%
{\sum_{j=1}^{L}\exp\!\left(-\frac{(j - \mu_k)^2}{2\sigma_k^2}\right)},
\end{equation}
where $P(i)$ denotes the predicted attention probability for token position $i$ in a sequence of length $L$.  
Each token position corresponds to a code token aligned with an AST node, thus representing a specific syntactic or semantic unit in the source code.  
$w^{(k)}$ is the normalized weight of the $k$-th mode, and the denominator ensures each Gaussian is properly normalized over all token positions.  
Smaller $\sigma_k$ values produce sharper, concentrated peaks representing focused reading, whereas larger $\sigma_k$ values yield broader distributions that capture peripheral attention.  
The resulting mixture $P(i)$ forms a smooth, interpretable, and differentiable attention distribution that aligns with empirical human fixation patterns and supports end-to-end optimization. The resulting distribution (P(i)) serves as the human-aligned attention prior used in the subsequent causal-aware redistribution stage (Section \ref{sec:redistribution}).

\subsection{Causal-Aware Attention Redistribution}\label{sec:redistribution}

Integrating human-guided attention into decoder-only transformers predicted by the EyeLayer requires preserving their causal autoregressive dependency.  
Unlike encoder-based models that permit bidirectional attention, decoder-only architectures must maintain strict left-to-right information flow so that each token prediction depends only on preceding context.  
Directly modifying attention weights or masks would break this constraint and disrupt key–value caching during generation.  
To address this, we implement \textit{causal-aware redistribution}, which injects human-aligned guidance through residual perturbations of hidden states rather than altering attention masks.  
The perturbation is shaped by the Gaussian attention distribution predicted by the EyeLayer, enabling soft alignment toward human-attended regions while fully preserving causality.  
The mechanism proceeds in three stages: (1) low-rank transformation for compact perturbation generation, (2) attention-guided weighting for cognitively informed modulation, and (3) adaptive gating for dynamic integration control.

\subsubsection{Low-Rank Transformation}

To prevent overfitting on limited eye-tracking data, perturbations are generated through a low-rank bottleneck.  
Given hidden states $\mathbf{H} \in \mathbb{R}^{B \times L \times d}$ from a target transformer layer, we first down-project and then reconstruct the representations:
\begin{align}
\mathbf{Z} &= \text{ReLU}(\mathbf{H} \mathbf{W}_{\text{down}}), \\
\Delta\mathbf{H}_{\text{base}} &= \mathbf{Z} \mathbf{W}_{\text{up}},
\end{align}
where $\mathbf{W}_{\text{down}} \in \mathbb{R}^{d \times r}$ and $\mathbf{W}_{\text{up}} \in \mathbb{R}^{r \times d}$ are learnable projections with rank $r \ll d$.  
This factorization requires only $2dr$ parameters instead of $d^2$, providing a $64\times$ reduction when $r=16$ for $d=2048$, while retaining sufficient representational capacity.

\subsubsection{Attention-Guided Weighting}

The perturbation is reweighted according to the predicted Gaussian attention distribution, emphasizing regions that align with human gaze.  
For each sample $b$ and token position $i$, we compute:
\begin{equation}
\tilde{\Delta}\mathbf{H}_{b,i} = \lambda\, P_b(i)\, \Delta\mathbf{H}_{\text{base},b,i} \odot A_i,
\end{equation}
where $P_b(i)$ denotes the mixture-based attention probability at position $i$, $\lambda$ is a learnable scaling coefficient, $A_i \in \{0,1\}$ marks valid token positions, and $\odot$ represents element-wise multiplication.  
Since redistribution operates on hidden representations rather than attention masks, causal self-attention remains intact: each token still attends only to past positions ($j \le i$), while its representation is softly modulated toward human-attended regions.  
Gradient clipping is applied to ensure numerical stability.

\subsubsection{Adaptive Gating and Integration}

Finally, an adaptive highway gate controls the strength of human-guided perturbation for each sample.  
A scalar gate value $g_b \in [0, g_{\max}]$ is computed as:
\begin{equation}
g_b = g_{\max}\, \sigma(\text{MLP}([\bar{\mathbf{h}}_b; \mathbf{f}_b])),
\end{equation}
where $\bar{\mathbf{h}}_b = \tfrac{1}{L}\sum_{i=1}^{L} \mathbf{H}_{b,i}$ is the mean-pooled hidden state (layer-normalized before concatenation), $\mathbf{f}_b$ encodes global statistics of the attention distribution (e.g., entropy, maximum probability, and in the multimodal case, mode count and weight entropy), and $\sigma(\cdot)$ is the sigmoid activation.  
The MLP is initialized with a negative bias to encourage conservative gating during early training.  
The final hidden states are obtained via residual integration:
\begin{equation}
\mathbf{H}'_{b,i} = \mathbf{H}_{b,i} + \alpha\, g_b\, \tilde{\Delta}\mathbf{H}_{b,i},
\end{equation}
where $\alpha$ is a global scaling constant.  
When $g_b$ is small, the EyeLayer exerts minimal influence; as $g_b$ increases, stronger redistribution occurs, enabling adaptive incorporation of human attention signals while preserving the model’s pretrained representations.

\subsection{Model Integration}\label{sec:model_integration}

\begin{figure}[htbp]
    \centering
    \includegraphics[width=0.45\textwidth]{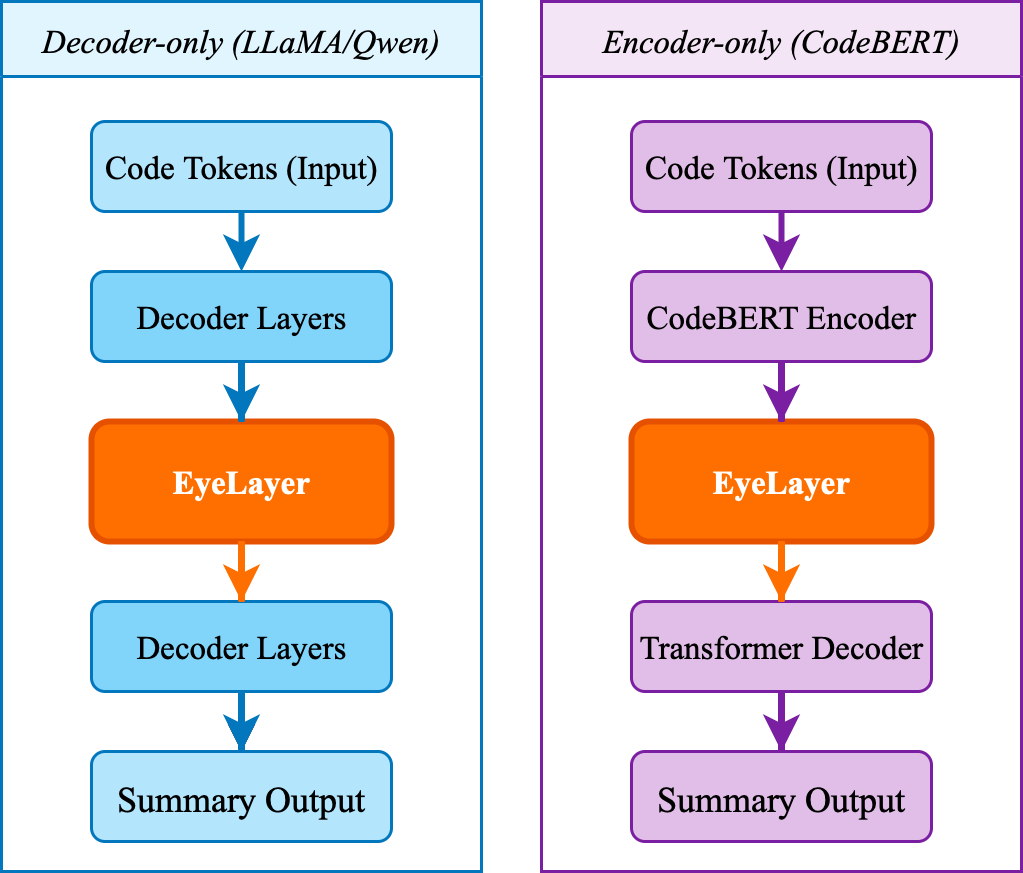}
    \caption{Integration of the EyeLayer into transformer architectures for code summarization. Note that since CodeBERT is an encoder-only model, an auxiliary decoder is attached for sequence generation in the code summarization task.}
    \Description{The figure shows how the EyeLayer is integrated into decoder-only and encoder-only transformer architectures for code summarization.}
    \label{fig:integration}
\end{figure}

The Multimodal EyeLayer integrates with transformer architectures through strategies that respect their information flow, as shown in Figure~\ref{fig:integration}.

\textbf{Decoder-Only Models (LLaMA, Qwen).}
For autoregressive decoder-only architectures, the EyeLayer is injected at an intermediate transformer layer. During forward propagation, when the base model reaches the target layer, the hook intercepts hidden states $\mathbf{H}$, applies the EyeLayer transformation, and returns enhanced representations $\mathbf{H}'$ to subsequent layers. The predicted distribution $P(i)$ guides causal-aware attention redistribution (Section~\ref{sec:redistribution}), which enforces that token $i$ only attends to positions $j \leq i$ and preserves the decoder’s generation order.

\textbf{Encoder-Only Models (CodeBERT).}
For encoder-only architectures, the EyeLayer operates after CodeBERT and before the auxiliary decoder. CodeBERT processes the input code to produce contextualized hidden states $\mathbf{H}_{\text{enc}}$, which are pooled to obtain a global code embedding that drives gating and mode prediction. The resulting $P(i)$ modulates $\mathbf{H}_{\text{enc}}$ via a non-causal low-rank perturbation over token positions; causal masking is not applied because the encoder is bidirectional. The decoder then cross-attends to the modulated encoder representations enriched with human-aligned attention priors.

\subsection{Joint Training}

After integrating the EyeLayer into the model architecture, we jointly train the system on the primary code summarization and auxiliary eye-tracking alignment tasks.  
This joint learning setup allows the model to balance large-scale textual supervision with sparse but cognitively grounded human signals.  
Formally, the overall objective combines a generation loss $\mathcal{L}_{\text{gen}}$ and an auxiliary alignment loss $\mathcal{L}_{\text{align}}$ (defined in Section~\ref{sec:alignment_loss}):
\begin{equation}
\mathcal{L}_{\text{total}} = \mathcal{L}_{\text{gen}} + \lambda_{\text{align}}\,\mathcal{L}_{\text{align}},
\end{equation}
where $\lambda_{\text{align}}$ is a small weighting coefficient that ensures the alignment supervision acts as a regularizer rather than dominating optimization.

\subsubsection{Interleaved Training Schedule}

Because the two datasets differ greatly in scale, with tens of thousands of code-summary pairs and only hundreds of eye-tracking samples, we adopt an interleaved training schedule to maintain stability.  
During each epoch, the model primarily trains on the summarization dataset, updating parameters with $\mathcal{L}_{\text{gen}}$ at every step.  
Every $K$ steps (typically $K=200$), a batch from the eye-tracking dataset is inserted, and EyeLayer is optimized jointly on $\mathcal{L}_{\text{total}}$ with gradient conflict handling described in Section~\ref{sec:pcgrad}.  
At the end of each epoch, we conduct a dedicated alignment sweep over the entire eye-tracking dataset while freezing the base model parameters, updating only the EyeLayer components.  
This two-phase schedule maintains consistent exposure to the generation objective and provides sufficient gradient signal for the EyeLayer through dedicated alignment phases, preventing the alignment objective from being overshadowed by the main summarization task.

\subsubsection{Projecting Conflicting Gradients (PCGrad)}\label{sec:pcgrad}

Multi-task optimization often leads to conflicting gradient directions between objectives.  
In our setting, the EyeLayer parameters are influenced by both $\mathcal{L}_{\text{gen}}$ and $\mathcal{L}_{\text{align}}$, which may occasionally compete.  
To reconcile these objectives, we employ \textbf{Projecting Conflicting Gradients (PCGrad)}~\cite{PCGrad}, which detects negative cosine similarity between task gradients and removes the conflicting component through orthogonal projection.  
When gradients are aligned, both signals are preserved; when they diverge, PCGrad adjusts each gradient to retain only the non-conflicting directions.  
The final parameter update uses the mean of the projected gradients, ensuring that human-guided supervision complements rather than disrupts the main learning objective.

\section{Experimental Setup}\label{sec:experiment_setup}

This section details the experimental configuration used to evaluate the proposed Multimodal Gaussian EyeLayer. We describe (1) dataset construction for both code summarization and eye-tracking supervision, (2) models and training infrastructure, and (3) evaluation metrics for summarization quality and human attention alignment. These components collectively establish the framework for answering the research questions presented in Section~\ref{sec:results}.

\subsection{Datasets}

\textbf{Code Summarization Dataset.}  
We use a subset of CodeXGLUE~\cite{codexglue}, derived from CodeSearchNet-Java, as the primary supervision source. To reduce training cost while preserving data diversity, we sample 10\% of the corpus, yielding 16,492 training pairs, 518 validation pairs, and 1,095 test pairs. Each instance consists of a Java method paired with its corresponding docstring summary extracted from open-source repositories. We adopt Java docstrings as ground-truth summaries following common practice in prior code summarisation research, as these docstrings are manually written by developers and widely used as API-level documentation.

\textbf{Eye-Tracking Dataset.}  
We adopt the EyeTrans corpus~\cite{Eyetrans} for human attention supervision. The corpus involves fixation data from 27 programmers performing code summarization tasks. Each data point corresponds to a unique (developer, method) pair, covering 64 unique functions across diverse Java projects. We obtain 625 annotated samples with fixation sequences aligned to AST nodes. These samples provide sparse but cognitively grounded supervision for guiding attention redistribution.

\subsection{Models and Training Infrastructure}

We evaluate our Multimodal Gaussian EyeLayer across three representative transformer architectures spanning different scales and designs: LLaMA3.2-1B and LLaMA3.2-3B (decoder-only instruction-tuned models), Qwen3-1.7B and Qwen3-4B (decoder-only base model), and CodeBERT (encoder-only code model). Training and evaluation are conducted on a single NVIDIA L40S GPU (45GB VRAM), confirming that our approach remains computationally efficient while effectively incorporating human attention guidance.

\subsection{Evaluation Metrics}

\subsubsection{Code Summarization Metrics}

We evaluate generation quality using four widely adopted metrics:
\begin{itemize}
\item \textbf{BLEU}~\cite{papineniBleuMethodAutomatic2002a}: Computes modified n\text{-}gram precision with a brevity penalty to quantify lexical overlap with references.
\item \textbf{ROUGE-L}~\cite{linROUGEPackageAutomatic2004}: Measures F1 based on the longest common subsequence, reflecting sequence-level similarity.
\item \textbf{METEOR}~\cite{banerjeeMETEORAutomaticMetric2005}: Aligns words using exact, stem, and synonym matches with fragmentation penalties, emphasizing recall and paraphrase recognition.
\item \textbf{BERTScore}~\cite{zhangBERTScoreEvaluatingText2020}: Computes contextual embedding similarity to assess semantic alignment between candidate and reference texts.
\end{itemize}

\subsubsection{Attention Alignment Metrics}
\label{sec:alignment_loss}

To align model-predicted attention with human fixation patterns while preserving multimodal diversity, we define:
\begin{equation}
\mathcal{L}_{\text{align}} = \mathcal{L}_{\text{match}} + \lambda_{\text{sep}} \mathcal{L}_{\text{MSP}},
\end{equation}
where $\mathcal{L}_{\text{match}}$ aligns each Gaussian mode with fixation data, and $\mathcal{L}_{\text{MSP}}$ enforces spatial separation among active modes.  
The matching term is computed as  
$\mathcal{L}_{\text{match}} = \sum_{k=1}^{K} \tilde{w}^{(k)} \sum_{t} \lambda_{t} \mathcal{L}_{t}^{(k)}$,  
where $t \in \{\text{CAL}, \text{SML}, \text{CR}, \text{AUP}\}$ and $\tilde{w}^{(k)}$ is the normalized mode weight.  
Here $w^{(k)}$, $\mu_k$, and $\sigma_k$ denote the weight, center, and spread of the $k$-th Gaussian;  
$P_k(i)$ is its normalized probability at token position $i$;  
$F(i)$ is the human fixation frequency; and $\epsilon$ is a small stability constant.

\textbf{Centroid Alignment (CAL).}  
$\mathcal{L}^{(k)}_{\text{CAL}}=\sqrt{(\mu_k-\mu_{\text{human}})^2+\epsilon}$,  
where $\mu_{\text{human}}$ is the empirical fixation centroid.  
This term aligns each predicted attention center $\mu_k$ with human focus regions.

\textbf{Spread Matching (SML).}  
$\mathcal{L}^{(k)}_{\text{SML}}=\sqrt{(\sigma_k-\sigma_{\text{target}})^2+\epsilon}$,  
where $\sigma_{\text{target}}$ represents the observed fixation spread.  
It ensures each mode captures realistic human attention breadth.

\textbf{Concentration Reward (CR).}  
$\mathcal{L}^{(k)}_{\text{CR}}=1-\big(\sum_{i\in\mathcal{W}}P_k(i)\big)^2$,  
where $\mathcal{W}$ is a local window around attended tokens.  
This rewards probability mass concentrated near human fixation areas.

\textbf{Anti-Uniform Penalty (AUP).}  
$\mathcal{L}^{(k)}_{\text{AUP}}=\max(0,\,c-D_{\mathrm{KL}}(U\Vert P_k))$,  
where $U(i)=1/L$ is the uniform baseline, $D_{\mathrm{KL}}$ is KL divergence,  
and $c$ is a small positive margin controlling the penalty strength.  
This term penalizes near-uniform distributions and promotes sharper attention focus.

\textbf{Mode Separation (MSP).}  
$\mathcal{L}_{\text{MSP}}=\!\sum_{k_1<k_2}\!s_{k_1k_2}\max(0,\,m-|\mu_{k_1}-\mu_{k_2}|)$,  
where $s_{k_1k_2}=\mathbb{I}[w^{(k_1)}>\tau]\mathbb{I}[w^{(k_2)}>\tau]$,  
$\tau$ is the activation threshold, and $m$ is the minimum distance between active mode centers.  
This term maintains spatial diversity across Gaussian components.

\section{Experimental Results and Analysis}\label{sec:results}

To evaluate the proposed Multimodal Gaussian EyeLayer, we address four research questions designed to quantify its effect on model performance, architectural behavior, and design components.

\begin{itemize}
\item \textbf{RQ1 – Does EyeLayer improve code summarization quality compared to standard supervised finetuning?}
\item \textbf{RQ2 – How does the position of the EyeLayer within the transformer stack influence performance?}
\item \textbf{RQ3 – How effectively does the EyeLayer generalize to encoder-only architectures?}
\item \textbf{RQ4 – How does EyeLayer multimodal design contribute to performance?}
\end{itemize}

\subsection{RQ1: Effectiveness Compared to SFT}


\begin{table*}[t]
\centering
\caption{Performance comparison of baseline models and models with EyeLayer integration.
Values in parentheses denote absolute improvement over the SFT baseline.}
\Description{The table compares BLEU-4, ROUGE-L, METEOR, and BERTScore for SFT baselines versus EyeLayer-augmented models, reporting absolute improvements in parentheses.}
\label{tab:rq1_baseline_comparison}
\begin{tabular}{l|cccc}
\toprule
\textbf{Model} & \textbf{BLEU-4} & \textbf{ROUGE-L} & \textbf{METEOR} & \textbf{BERTScore} \\
\midrule
\rowcolor{white}
Llama3.2-1B & 14.31 & 22.12 & 27.45 & 87.55 \\
\rowcolor{gray!10}
Llama3.2-1B + EyeLayer & 16.18 (\textbf{+1.87}) & 23.51 (\textbf{+1.39}) & 29.33 (\textbf{+1.88}) & 88.51 (\textbf{+0.96}) \\
\midrule
\rowcolor{white}
Llama3.2-3B & 15.64 & 24.57 & 29.83 & 88.29 \\
\rowcolor{gray!10}
Llama3.2-3B + EyeLayer & 16.86 (\textbf{+1.22}) & 25.25 (\textbf{+0.68}) & 31.04 (\textbf{+1.21}) & 88.72 (\textbf{+0.43}) \\
\midrule
\midrule
\rowcolor{white}
Qwen3-1.7B & 13.36 & 21.39 & 26.60 & 86.04 \\
\rowcolor{gray!10}
Qwen3-1.7B + EyeLayer & 15.12 (\textbf{+1.76}) & 26.67 (\textbf{+5.28}) & 32.03 (\textbf{+5.43}) & 86.38 (\textbf{+0.34}) \\
\midrule
\rowcolor{white}
Qwen3-4B & 15.24 & 23.73 & 29.45 & 85.87 \\
\rowcolor{gray!10}
Qwen3-4B + EyeLayer & 17.22 (\textbf{+1.98}) & 25.30 (\textbf{+1.57}) & 31.31 (\textbf{+1.86}) & 86.27 (\textbf{+0.40}) \\
\bottomrule
\end{tabular}

\end{table*}
RQ1 investigates whether integrating the proposed EyeLayer improves code summarization quality compared to standard supervised finetuning (SFT) without eye-tracking guidance. We evaluate four representative models: instruction-tuned (Llama3.2-1B/3B) and base (Qwen3-1.7B/4B).

As shown in Table~\ref{tab:rq1_baseline_comparison}, integrating EyeLayer leads to consistent gains across all models and evaluation metrics. Improvements appear in both lexical metrics (BLEU, ROUGE, METEOR) and semantic similarity (BERTScore), suggesting that cognitively inspired attention cues can guide the model toward more functionally meaningful code regions. For instruction-tuned models (Llama3.2-instruct), EyeLayer yields steady gains, particularly for the 1B model (+1.8 BLEU-4 / +1.9 METEOR). For base models(Qwen3), the improvement is larger in absolute terms, particularly for Qwen3-1.7B (ROUGE-L: +5.28, METEOR: +5.43), which indicates that models lacking instruction-level supervision may benefit more from additional attention prior.

The performance improvement suggests that EyeLayer subtly guides intermediate attention toward critical code regions that typically attract human gaze, thereby improving the quality of generated summary. The relatively larger gains observed in smaller models imply that supervision from the eye-tracking corpus provides a more informative inductive signal when model capacity and learned abstractions are limited. Larger models which already develop rich internal attention patterns, exhibit smaller yet consistent benefits. These observations collectively point to EyeLayer as a light but effective cognitive guidance mechanism, offering additional structure to models operating under supervision.

\begin{figure}[t]
    \centering
    \includegraphics[width=0.95\linewidth]{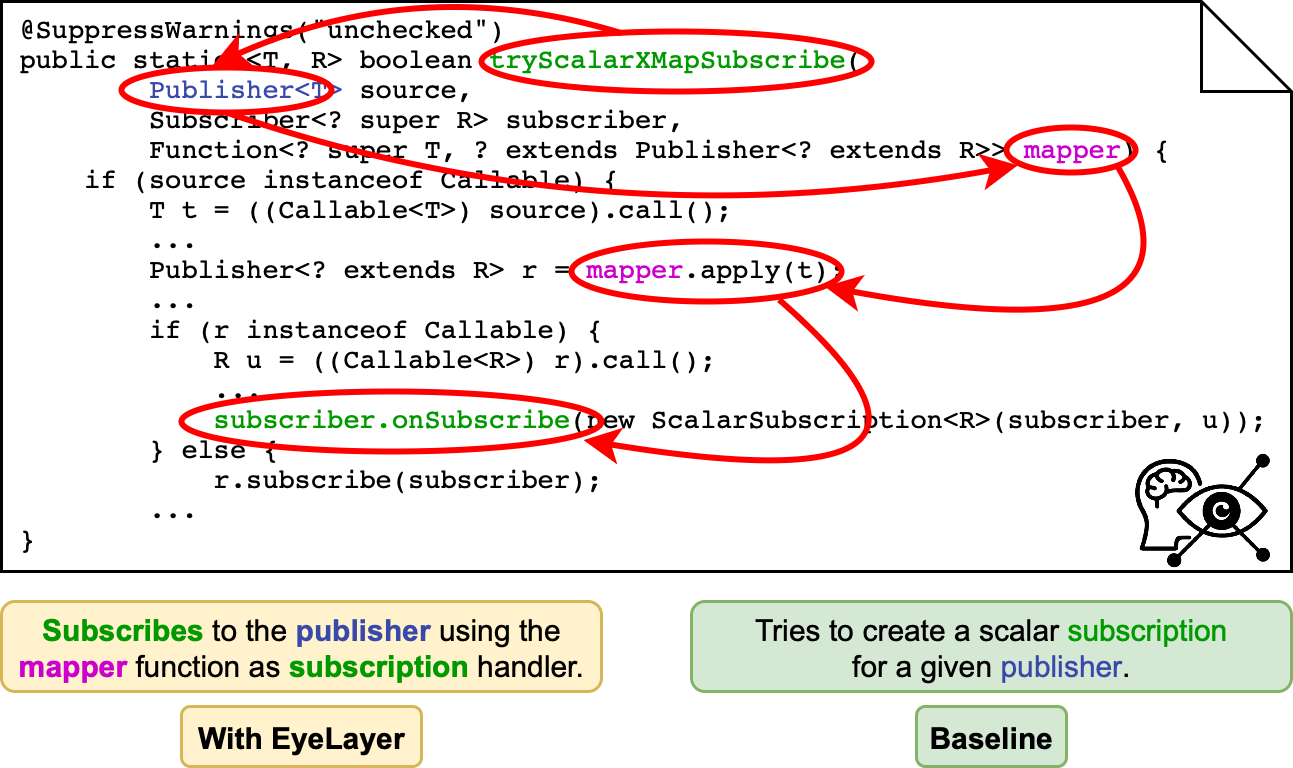}
    \caption{
    Example from CodeXGLUE illustrating EyeLayer’s improvement over the baseline. 
    Depict the inferred gaze-inspired attention across semantically related code regions.
    }
    \Description{The figure presents a case study comparing baseline and EyeLayer summaries and visualizes stronger attention on semantically important code regions with EyeLayer.}
    \label{fig:rq1_case_study}
\end{figure}

Figure~\ref{fig:rq1_case_study} illustrates a representative example that demonstrates how EyeLayer enhances the generated summary. The baseline output, \textit{“Tries to create a scalar subscription for a given publisher,”} captures only surface lexical cues, whereas EyeLayer produces a more accurate behavioral description, \textit{“Subscribes to the publisher using the mapper function as subscription handler.”} 
Compared to the baseline, EyeLayer places stronger focus on the method declaration and variable declarations, which are semantically critical regions for capturing functional intent. This pattern resonates with the human attention dynamics reported by Karas et~al.~\cite{karas2024tale}, where programmers most frequently alternate their gaze between method declarations and variable declarations during code comprehension. 
The correspondence suggests that EyeLayer internalizes similar focus tendencies without explicit gaze supervision during inference, enabling the model to generalize cognitive attention patterns that guide summarization toward semantically informative code regions.

\textbf{RQ1 Summary.}
EyeLayer consistently improves summarization across all models, with larger gains in smaller or less supervised settings, showing that lightweight cognitive cues enhance semantic focus in code comprehension.

\subsection{RQ2: Effect of EyeLayer Insertion Position}

\begin{figure}[htbp]
  \centering
  \begin{subfigure}[b]{0.22\textwidth}
    \includegraphics[width=\textwidth]{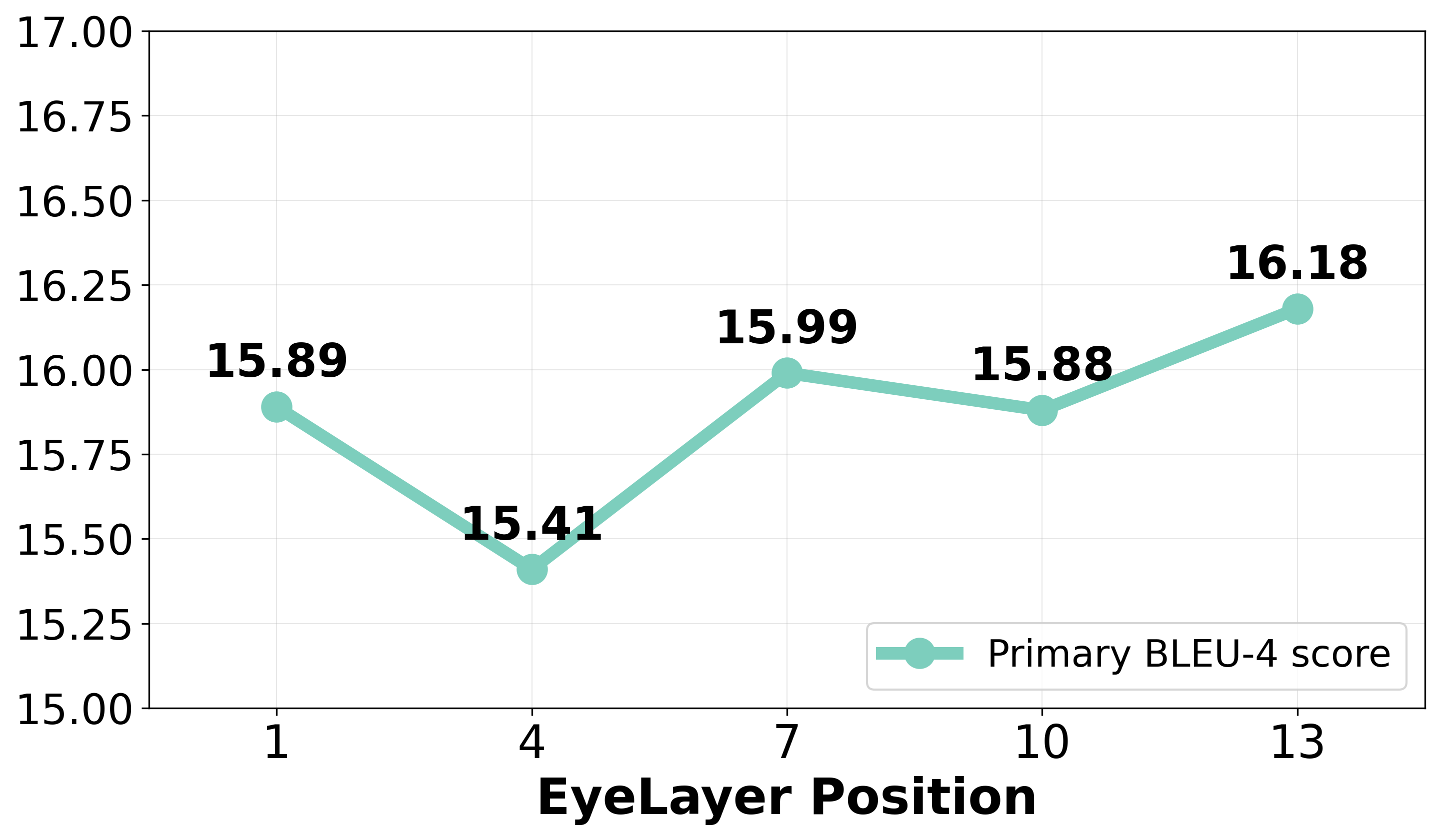}
    \caption{BLEU-4}
  \end{subfigure}
  \begin{subfigure}[b]{0.22\textwidth}
    \includegraphics[width=\textwidth]{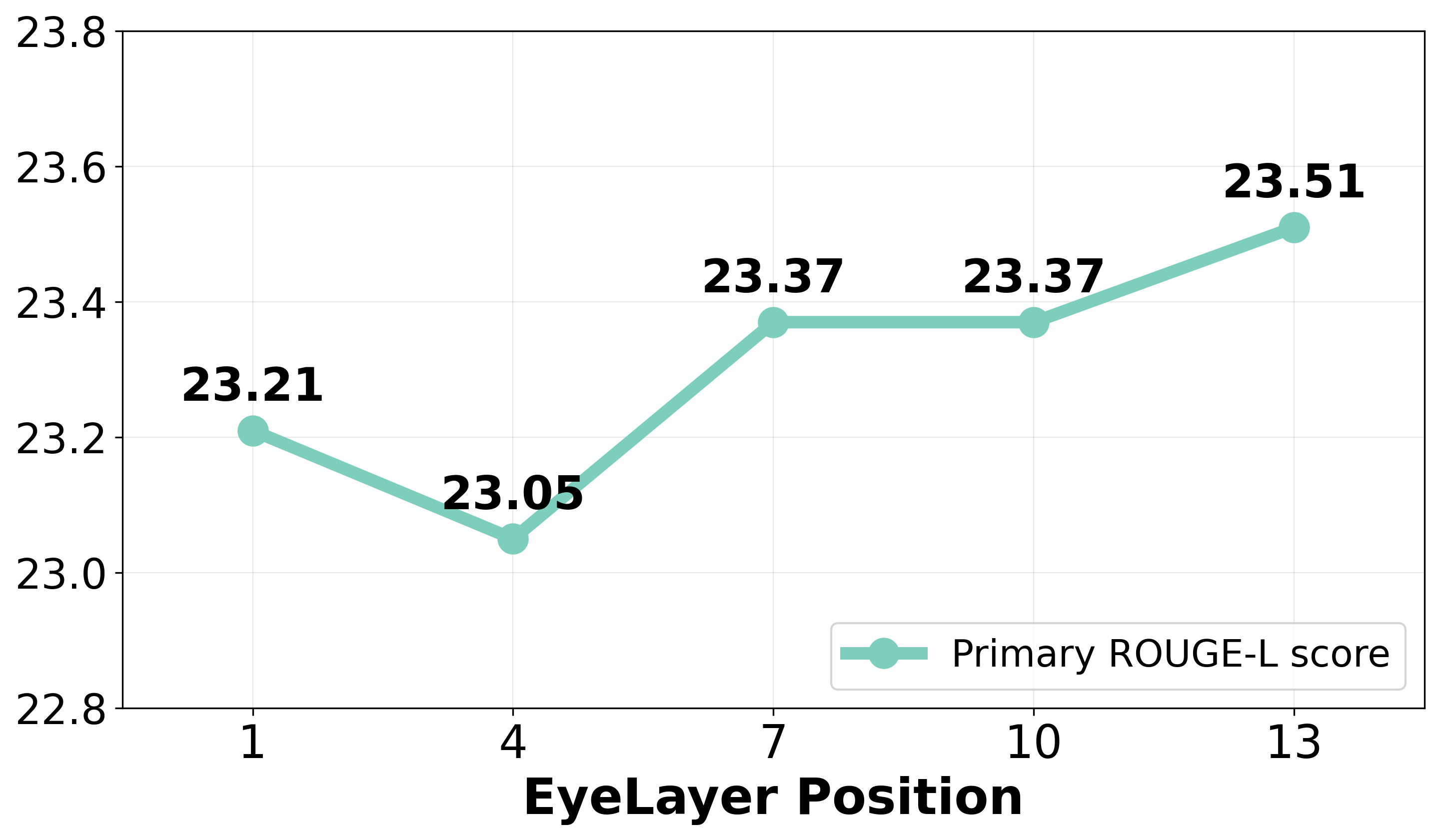}
    \caption{ROUGE-L}
  \end{subfigure}

  \begin{subfigure}[b]{0.22\textwidth}
    \includegraphics[width=\textwidth]{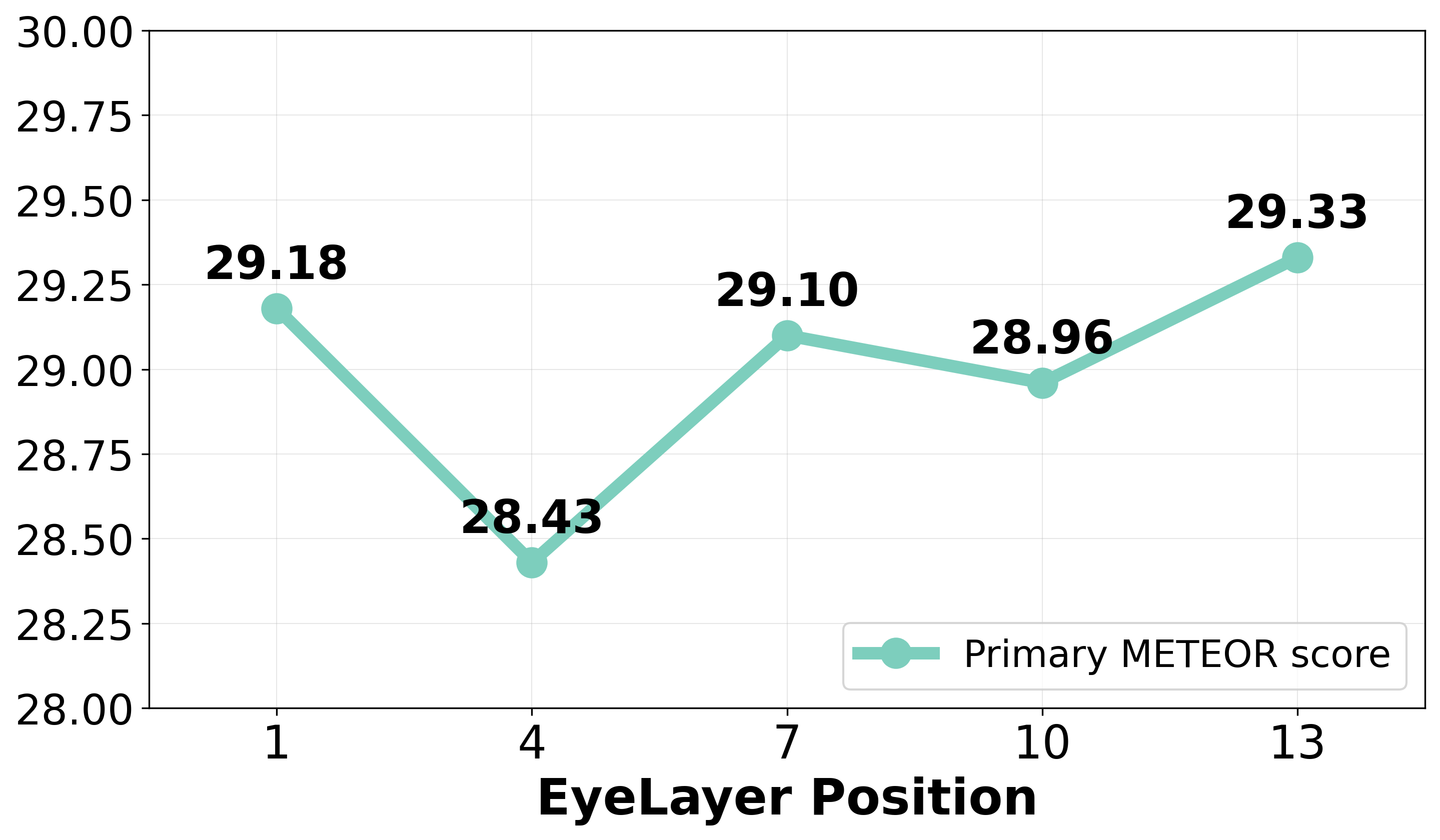}
    \caption{METEOR}
  \end{subfigure}
  \begin{subfigure}[b]{0.22\textwidth}
    \includegraphics[width=\textwidth]{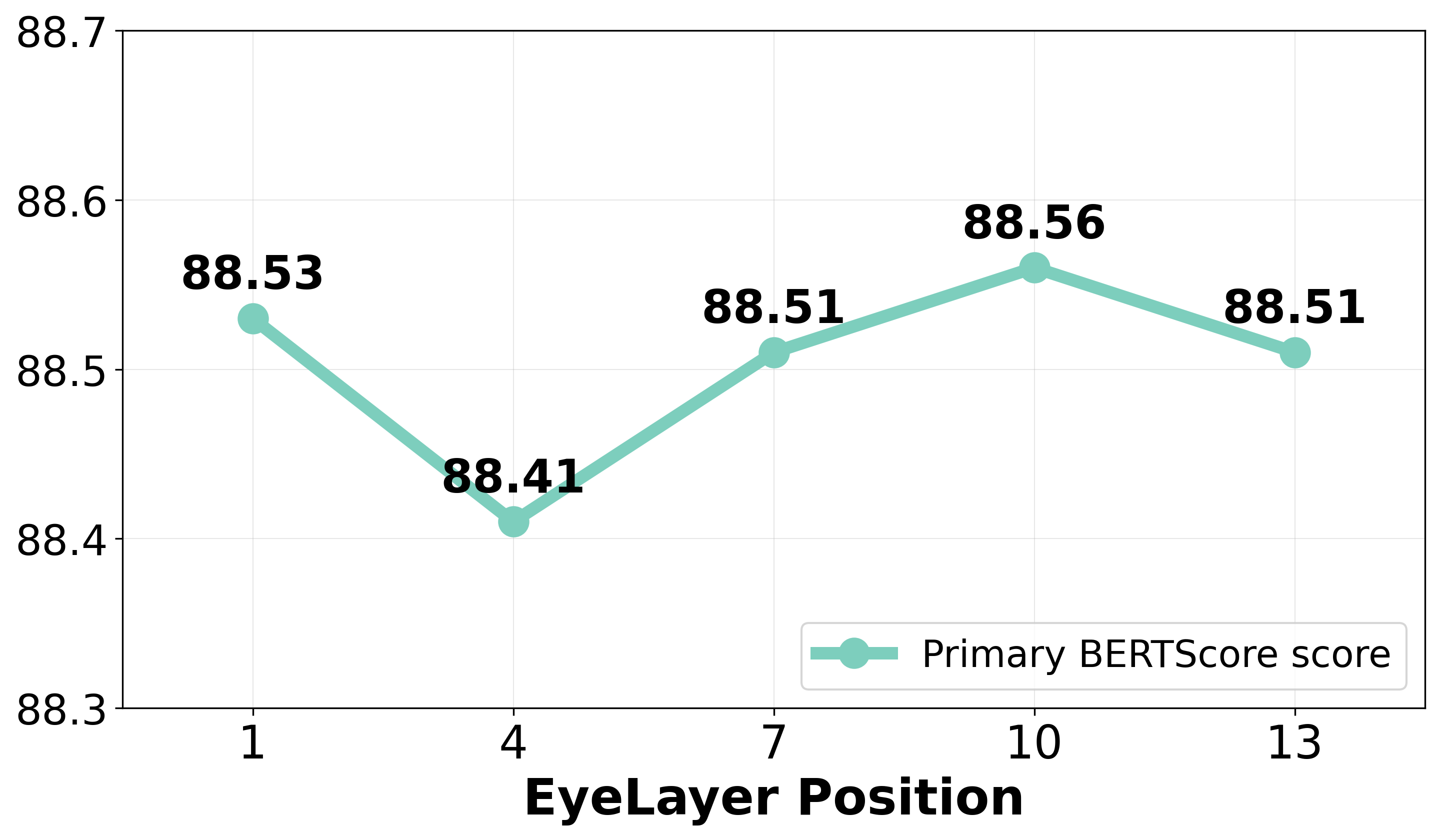}
    \caption{BERTScore}
  \end{subfigure}

    \caption{Performance of Llama3.2-1B-Instruct when the EyeLayer is inserted at different transformer layers.}
    \Description{The figure plots four evaluation metrics for Llama3.2-1B-Instruct when EyeLayer is inserted at different transformer layers.}
    \label{fig:layer-position-grid}
\end{figure}

We investigate how integration depth affects performance by inserting the EyeLayer into different transformer layers of Llama3.2-1B-Instruct (16 layers). Figure~\ref{fig:layer-position-grid} shows the different metric trends across positions.

Two clear patterns emerge: (1) performance improves toward deeper layers and peaks at layer13, and (2) a temporary drop appears around layer~4. This trend aligns with the hierarchical roles of transformer layers~\cite{merulloTalkingHeadsUnderstanding2025a,skeanLayerLayerUncovering2025}. Early layers capture lexical and syntactic features, middle layers integrate contextual semantics, and later-middle layers refine coherent representations for generation~\cite{fartaleDisentanglingRecallReasoning2025}. The degradation at layer~4 likely reflects interference with unstable intermediate encodings, as this stage is still reorganizing shallow features into higher-level structures~\cite{zhangInvestigatingLayerImportance2024}. At layer~13, semantic representations are largely formed yet remain adaptable. Injecting human attention priors here allows modulation of semantic focus without disrupting earlier composition, enhancing alignment with meaningful program structures~\cite{skeanLayerLayerUncovering2025,merulloTalkingHeadsUnderstanding2025a}.

Overall, these results highlight that the integration of cognitive priors depends strongly on the model’s representational stage, with later-middle layers providing the best balance between semantic completeness and flexibility~\cite{zhangInvestigatingLayerImportance2024,fartaleDisentanglingRecallReasoning2025}.

\textbf{RQ2 Summary.}
Performance peaks at later-middle layers, where semantic representations are mature yet flexible, indicating that cognitive priors are most effective after semantic integration but before generation.

\subsection{RQ3: Generalization to Encoder-Only Architectures}

Building on the results from decoder-only models (RQ1) and the optimal integration depth analysis (RQ2), RQ3 examines whether EyeLayer generalizes to encoder-only architectures, which differ fundamentally in information flow and attention dynamics. We evaluate this transferability using CodeBERT with the encoder-side integration strategy described in Section~\ref{sec:model_integration}. The results are summarized in Table~\ref{tab:rq3_codebert}.

\begin{table*}[t]
\centering
\caption{CodeBERT performance with and without EyeLayer integration.}
\Description{The table reports CodeBERT performance with and without EyeLayer using BLEU-4, ROUGE-L, METEOR, and BERTScore.}
\label{tab:rq3_codebert}
\begin{tabular}{l|cccc}
\toprule
\textbf{Model} & \textbf{BLEU-4} & \textbf{ROUGE-L} & \textbf{METEOR} & \textbf{BERTScore} \\
\midrule
\rowcolor{white}
CodeBERT & 14.35 & 29.16 & 21.87 & 87.69 \\
\rowcolor{gray!10}
CodeBERT + EyeLayer & 15.39 (\textbf{+1.04}) & 30.70 (\textbf{+1.54}) & 23.70 (\textbf{+1.83}) & 88.30 (\textbf{+0.61}) \\
\bottomrule
\end{tabular}
\end{table*}

EyeLayer maintains consistent improvements across all metrics, despite the architectural shift from decoder-only to encoder-only models. The largest gain appears in METEOR (+1.83), indicating enhanced semantic alignment and paraphrase understanding, both of which rely on holistic code comprehension. The bidirectional encoder benefits from modeling human-like focus over the entire code context without causal masking, explaining its strong performance on semantic metrics.

These results suggest that human attention patterns encode architecture-invariant cues of semantic importance. Regardless of whether information is processed autoregressively or bidirectionally, guiding attention toward regions that typically attract human gaze helps redistribute representational focus more effectively. The multimodal Gaussian formulation accommodates these differences without architectural redesign, demonstrating EyeLayer’s flexibility and generalizability as a cognitively grounded attention module.

\textbf{RQ3 Summary.}
Results on CodeBERT indicate that EyeLayer can be applied to an encoder-only architecture, suggesting that human attention patterns may offer architecture-agnostic cues for guiding semantic focus.


\subsection{RQ4: Ablation Study on Multimodal Design}

\begin{table*}[t]
\centering
\caption{Ablation study comparing single-mode and multimodal EyeLayer designs on Llama3.2-1B.}
\Description{The table shows an ablation comparing baseline, single-mode, and multimodal EyeLayer configurations on Llama3.2-1B across four summarization metrics.}
\label{tab:rq4_ablation}
\begin{tabular}{l|cccc}
\toprule
\textbf{Configuration} & \textbf{BLEU-4} & \textbf{ROUGE-L} & \textbf{METEOR} & \textbf{BERTScore} \\
\midrule
\rowcolor{white}
Baseline (SFT) & 14.31 & 22.12 & 27.45 & 87.55 \\
\rowcolor{gray!10}
Single-mode (Early) & 14.30 & 21.64 & 27.55 & 88.13 \\
\rowcolor{white}
Single-mode (Late) & 14.63 & 20.82 & 26.10 & 88.26 \\
\rowcolor{gray!10}
Multimodal (Late) & \textbf{16.18} & \textbf{23.51} & \textbf{29.33} & \textbf{88.51} \\
\bottomrule
\end{tabular}
\end{table*}

RQ4 investigates whether the multimodal Gaussian design, which models human attention through multiple distinct modes, provides advantages over simpler single mode alternatives.

To isolate the multimodal design’s contribution, we implement a simplified EyeLayer variant that predicts attention using a single Gaussian distribution rather than a mixture. This architecture removes the sparse gating network and mode-specific prediction heads, and instead directly predicts the global centroid $\mu$ and spread $\sigma$ from the code-level embedding, while retaining all other components, including low-rank perturbation, attention-guided weighting, and adaptive highway gating.
We evaluate single-mode EyeLayer at early and late layer positions on Llama3.2-1B-instruct.

Table~\ref{tab:rq4_ablation} shows that multimodal EyeLayer consistently outperforms single-mode variants across all metrics. Single-mode configurations show limited improvements over baseline, with early layer achieving minimal gains and late layer showing inconsistent performance. In contrast, multimodal EyeLayer delivers substantial improvements. For example, at layer 13, multimodal design achieves BLEU-4: 16.18 versus single-mode's 14.63 (+1.55), and METEOR: 29.33 versus 26.10 (+3.23), demonstrating clear advantages of modeling multiple attention modes.

The results support our hypothesis that human attention during code comprehension cannot be captured by a single Gaussian. A single-mode design can only represent one attention region, which forces a trade-off between narrow focus (small $\sigma$) and broad coverage (large $\sigma$), and thus fails to model multiple distinct areas of interest in complex functions. In contrast, the multimodal design enables sparse mode selection, where the gating network activates 1–3 modes adaptively based on code complexity. This allows the model to compose multiple attention patterns, such as scanning function signatures, following control flow, and inspecting implementation details. The substantial performance gains indicate that modeling diverse attention modes improves cognitive fidelity and justifies the added architectural complexity.

\textbf{RQ4 Summary.}
Multimodal Gaussian design outperforms single-mode variants, demonstrating that modeling multiple attention modes better captures human gaze diversity and yields stronger semantic alignment.
\section{Threats to Validity}\label{sec:threats}
There are several threats to the validity of our study.

First, the eye-tracking supervision is limited in scale and collected from a small group of professional developers performing Java code comprehension tasks. This may introduce biases related to individual reading strategies or experience levels and may limit generalization to other developer populations. However, EyeLayer learns probabilistic attention priors rather than developer-specific gaze patterns, which may reduce this risk.

Second, our experiments focus exclusively on Java code, which may limit generalization to languages with different syntactic structures or paradigms. Prior work suggests that core code comprehension strategies are largely shared across languages, but further validation on additional languages is required.

Third, our evaluation relies on automatic metrics that may not fully capture human-perceived summary quality or real-world usefulness. We mitigate this threat by using multiple complementary metrics (BLEU, ROUGE-L, METEOR, and BERTScore), evaluating across both decoder-only and encoder-only architectures, and conducting qualitative analysis on representative examples to assess semantic coverage and faithfulness.

All experiments were conducted with fixed random seeds to ensure reproducibility and reduce stochastic variance.
\section{Discussion and Future Work}\label{sec:discussion}
\textbf{Scaling Eye-Tracking Supervision.} Our results show that 625 sparse eye-tracking samples provide consistent benefits, suggesting that scaling supervision through data augmentation or large-scale collection could further improve performance. Richer supervision would enable more expressive EyeLayer architectures capturing finer-grained attention patterns.

\textbf{Richer Cognitive Signals.} Our approach uses only static fixation—aggregated attention intensity. Eye-tracking can contain additional information: saccade patterns (revealing information-seeking strategies), and attention switches (capturing dynamic shifts in cognitive focus). Incorporating these temporal and sequential signals has the potential to provide richer supervision.

\textbf{Generalization to Software Engineering Tasks.} While we focus on code summarization, many SE tasks fundamentally involve code comprehension: bug localization, code review, and program repair all require identifying semantically important regions. Human attention patterns could transfer across tasks as developers employ similar cognitive strategies regardless of end goal. EyeLayer's effectiveness across both decoder-only and encoder-only architectures demonstrates its flexibility for integration into diverse models. Future work could investigate whether attention patterns from code summarization tasks can transfer to other SE contexts, or whether collecting task-specific eye-tracking data yields stronger supervision signals.

\textbf{Broader Implications.} Beyond performance improvements, EyeLayer demonstrates grounding neural models in human cognitive processes rather than purely data-driven learning. This approach could enable more interpretable AI systems where models attend to code for reasons aligned with human reasoning, facilitating developer trust and effective human-AI collaboration as code intelligence tools become ubiquitous in development workflows.
\section{Related Work}\label{sec:relatedwork}

This section situates our work at the intersection of human-centered AI and automatic code summarization. We first review research that integrates cognitive and behavioral signals into software engineering models, emphasizing eye-tracking as a bridge between human and machine attention. We then discuss advances in code summarization, from transformer-based architectures to recent efforts incorporating human-like attention guidance.

\subsection{Human-centered AI for Software Engineering}\label{sec:related_work_hcai_se}

Human-centered AI for software engineering (SE) emphasizes aligning automated systems with human cognition and developer workflows.  
Empirical studies have shown that developers interact with AI assistants in complex ways: they often exhibit overconfidence while producing less secure code~\cite{perry2023ccs}, alternate between acceleration and exploration modes depending on task certainty~\cite{barke2023grounded}, and face persistent challenges in output validation and trust calibration~\cite{liang2024icse,fu2024tosem,lyu2025ase}.  
Recent theoretical frameworks further characterize trust as a dynamic and multi-dimensional construct~\cite{sabouri2025icse_trust,choudhuri2025icse}, underscoring the need for models that are cognitively transparent and behaviorally adaptive.

Beyond behavioral analysis, recent research has sought to directly model cognitive processes underlying code comprehension. Early eye-tracking studies revealed that developer gaze patterns reflect semantic understanding during program reading~\cite{rodegheroImprovingAutomatedSource2014b,paltenghiThinkingDeveloperComparing2021}. Building on this foundation, Bansal et al.~\cite{bansalModelingHumanAttention2023b} and Alakmeh et al.~\cite{alakmehPredictingCodeComprehension2024a} predicted human attention from code structure and integrated gaze information to enhance summarization models. More recently, EyeTrans~\cite{Eyetrans} and EyeMulator~\cite{zhangEyeMulatorImprovingCode2025} incorporated gaze data into Transformer architectures, achieving measurable performance gains. 

\textit{EyeLayer} extends this research direction by being among the first to incorporate human cognitive signals into large language models.
It leverages human attention as a transferable probabilistic prior, aiming for generalizable integration of human-like focus patterns across model architectures and tasks.

\subsection{Automatic Code Summarization}

The advent of large language models (LLMs) has catalyzed a paradigm shift in automatic code summarization, transitioning from traditional sequence-to-sequence architectures to transformer-based approaches that leverage extensive pre-training on code corpora.  
Early work such as Code2Seq~\cite{alon2019code2seq} and retrieval-augmented methods~\cite{zhang2020retrieval} demonstrated that structural program representations and example-based retrieval can significantly enhance summary quality.  
The establishment of benchmarks like CodeXGLUE~\cite{lu2021codexglue} standardized evaluation protocols and enabled systematic comparison across models and datasets.  
Building on these foundations, Shi et al.~\cite{shi2022evaluation} identified key factors influencing neural summarization performance, while Gao et al.~\cite{gao2023makes} and Fang et al.~\cite{fang2025enhanced} explored in-context and prompt-based learning to adapt general-purpose LLMs for code summarization.  
Empirical studies further revealed that moderately sized, fine-tuned models can rival or surpass much larger general-purpose LLMs when supervision effectively captures task semantics~\cite{sun2024source}, emphasizing the centrality of the fine-tuning process in code-oriented adaptation.

Recent work has focused on improving efficiency, robustness, and interpretability in LLM-based summarization~\cite{sun2024source}.  
Su et al.~\cite{su2024distilled} applied knowledge distillation to reduce computational costs, while Mastropaolo et al.~\cite{mastropaolo2024evaluating} proposed semantic-aware evaluation metrics to better assess summary fidelity.  
Virk et al.~\cite{virk2025calibration} exposed calibration deficiencies that undermine model reliability, and Mondal et al.~\cite{mondal2023robust} examined robustness to adversarial perturbations.  
Interpretability analyses further uncovered a persistent misalignment between model-generated attention and developer comprehension:  
Li et al.~\cite{li2024machines} showed that neural attention often diverges from code regions developers focus on, leading to summaries that are lexically fluent but semantically incomplete.  
This gap between surface-level correlations and true comprehension has motivated recent studies to augment fine-tuning with auxiliary behavioral cues such as eye-tracking, exemplified by EyeTrans~\cite{Eyetrans}, which guide transformer attention toward semantically salient regions.

\textit{EyeLayer} continues this trajectory by strengthening the supervised fine-tuning of LLM-based summarization. Rather than redesigning model architectures or relying on heavy supervision, it introduces lightweight cognitive priors into the fine-tuning pipeline to steer attention toward functionally important code regions.  

\section{Conclusion}\label{sec:conclusion}

This work demonstrates that human cognitive patterns captured through eye-tracking can effectively enhance LLM-based code summarization. We introduced EyeLayer, a lightweight attention-augmentation module that integrates sparse human attention signals into LLMs through Multimodal Gaussian Mixture Models, enabling models to learn how developers naturally focus on semantically critical code regions during comprehension. Our evaluation across five models spanning different scales and architectures shows consistent improvements, validating that human expertise provides complementary signals that enhance LLM capabilities beyond what standard supervised fine-tuning achieves. Our methodology establishes a framework for incorporating human cognitive processes into LLMs for code comprehension, contributing to the development of more capable and interpretable developer tools as software systems continue to grow in complexity.

\begin{acks}
We appreciate the constructive suggestions provided by the anonymous reviewers, which helped improve our manuscript.
\end{acks}

\bibliographystyle{ACM-Reference-Format}
\bibliography{references}

@misc{fengCodeBERTPretrainedModel2020,
  title = {{{CodeBERT}}: {{A}} Pre-Trained Model for Programming and Natural Languages},
  shorttitle = {{{CodeBERT}}},
  author = {Feng, Zhangyin and Guo, Daya and Tang, Duyu and Duan, Nan and Feng, Xiaocheng and Gong, Ming and Shou, Linjun and Qin, Bing and Liu, Ting and Jiang, Daxin and Zhou, Ming},
  year = 2020,
  month = sep,
  number = {arXiv:2002.08155},
  publisher = {arXiv},
  langid = {english}
}

@misc{grattafioriLlama3Herd2024a,
  title = {The Llama 3 Herd of Models},
  author = {Grattafiori, Aaron and Dubey, Abhimanyu and Jauhri, Abhinav and Pandey, Abhinav and Kadian, Abhishek and {Al-Dahle}, Ahmad and Letman, Aiesha and Mathur, Akhil and et al.},
  year = 2024,
  month = nov,
  number = {arXiv:2407.21783},
  publisher = {arXiv},
  doi = {10.48550/arXiv.2407.21783}
}

@misc{yangQwen3TechnicalReport2025,
  title = {Qwen3 Technical Report},
  author = {Yang, An and Li, Anfeng and Yang, Baosong and Zhang, Beichen and Hui, Binyuan and Zheng, Bo and Yu, Bowen and et al.},
  year = 2025,
  month = may,
  number = {arXiv:2505.09388},
  publisher = {arXiv},
  doi = {10.48550/arXiv.2505.09388},
  langid = {american}
}

@article{husain2019codesearchnet,
  title={{CodeSearchNet} challenge: Evaluating the state of semantic code search},
  author={Husain, Hamel and Wu, Ho-Hsiang and Gazit, Tiferet and Allamanis, Miltiadis and Brockschmidt, Marc},
  journal={arXiv preprint arXiv:1909.09436},
  year={2019}
}

@article{codexglue,
  author    = {Shuai Lu and
               Daya Guo and
               Shuo Ren and
               Junjie Huang and
               Alexey Svyatkovskiy and
               Ambrosio Blanco and
               Colin B. Clement and
               Dawn Drain and
               Daxin Jiang and
               Duyu Tang and
               Ge Li and
               Lidong Zhou and
               Linjun Shou and
               Long Zhou and
               Michele Tufano and
               Ming Gong and
               Ming Zhou and
               Nan Duan and
               Neel Sundaresan and
               Shao Kun Deng and
               Shengyu Fu and
               Shujie Liu},
  title     = {CodeXGLUE: {A} Machine Learning Benchmark Dataset for Code Understanding
               and Generation},
  journal   = {CoRR},
  volume    = {abs/2102.04664},
  year      = {2021}
}

@article{Eyetrans,
   title={EyeTrans: Merging Human and Machine Attention for Neural Code Summarization},
   volume={1},
   ISSN={2994-970X},
   url={http://dx.doi.org/10.1145/3643732},
   DOI={10.1145/3643732},
   number={FSE},
   journal={Proceedings of the ACM on Software Engineering},
   publisher={Association for Computing Machinery (ACM)},
   author={Zhang, Yifan and Li, Jiliang and Karas, Zachary and Bansal, Aakash and Li, Toby Jia-Jun and McMillan, Collin and Leach, Kevin and Huang, Yu},
   year={2024},
   month=jul, pages={115–136} }

@article{karas2024tale,
author = {Karas, Zachary and Bansal, Aakash and Zhang, Yifan and Li, Toby and McMillan, Collin and Huang, Yu},
title = {A Tale of Two Comprehensions? Analyzing Student Programmer Attention during Code Summarization},
year = {2024},
issue_date = {September 2024},
publisher = {Association for Computing Machinery},
address = {New York, NY, USA},
volume = {33},
number = {7},
issn = {1049-331X},
url = {https://doi.org/10.1145/3664808},
doi = {10.1145/3664808},
journal = {ACM Trans. Softw. Eng. Methodol.},
month = aug,
articleno = {193},
numpages = {37}
}

@inproceedings{ahmad2020summarization,
 author = {Ahmad, Wasi Uddin and Chakraborty, Saikat and Ray, Baishakhi and Chang, Kai-Wei},
 booktitle = {Proceedings of the 58th Annual Meeting of the Association for Computational Linguistics (ACL)},
 title = {A Transformer-based Approach for Source Code Summarization},
 year = {2020}
}

@misc{hou2024largelanguagemodelssoftware,
      title={Large Language Models for Software Engineering: A Systematic Literature Review}, 
      author={Xinyi Hou and Yanjie Zhao and Yue Liu and Zhou Yang and Kailong Wang and Li Li and Xiapu Luo and David Lo and John Grundy and Haoyu Wang},
      year={2024},
      eprint={2308.10620},
      archivePrefix={arXiv},
      primaryClass={cs.SE},
      url={https://arxiv.org/abs/2308.10620}, 
}

@INPROCEEDINGS{fan_llm4se,
  author={Fan, Angela and Gokkaya, Beliz and Harman, Mark and Lyubarskiy, Mitya and Sengupta, Shubho and Yoo, Shin and Zhang, Jie M.},
  booktitle={2023 IEEE/ACM International Conference on Software Engineering: Future of Software Engineering (ICSE-FoSE)}, 
  title={Large Language Models for Software Engineering: Survey and Open Problems}, 
  year={2023},
  volume={},
  number={},
  pages={31-53},
  doi={10.1109/ICSE-FoSE59343.2023.00008}}

@article{grabingerEyeTrackingSoftware2024,
  title = {On Eye Tracking in Software Engineering},
  author = {Grabinger, Lisa and Hauser, Florian and Wolff, Christian and Mottok, J{\"u}rgen},
  year = 2024,
  month = jul,
  journal = {SN Computer Science},
  volume = {5},
  number = {6},
  pages = {729},
  issn = {2661-8907},
  doi = {10.1007/s42979-024-03045-3},
  langid = {english}
}

@article{justTheoryReadingEye,
  title = {A Theory of Reading: {{From}} Eye Fixations to Comprehension},
  author = {Just, Marcel Adam and Carpenter, Patricia A},
  langid = {english}
}

@inproceedings{sharafiEyetrackingMetricsSoftware2015,
  title = {Eye-Tracking Metrics in Software Engineering},
  booktitle = {2015 {{Asia-Pacific Software Engineering Conference}} ({{APSEC}})},
  author = {Sharafi, Zohreh and Shaffer, Timothy and Sharif, Bonita and Gu{\'e}h{\'e}neuc, Yann-Ga{\"e}l},
  year = 2015,
  month = dec,
  pages = {96--103},
  issn = {1530-1362},
  doi = {10.1109/APSEC.2015.53},
  langid = {american}
}

@article{sharafiPracticalGuideConducting2020b,
  title = {A Practical Guide on Conducting Eye Tracking Studies in Software Engineering},
  author = {Sharafi, Zohreh and Sharif, Bonita and Gu{\'e}h{\'e}neuc, Yann-Ga{\"e}l and Begel, Andrew and Bednarik, Roman and Crosby, Martha},
  year = 2020,
  month = sep,
  journal = {Empirical Software Engineering},
  volume = {25},
  number = {5},
  pages = {3128--3174},
  publisher = {{Springer Science and Business Media LLC}},
  issn = {1382-3256, 1573-7616},
  doi = {10.1007/s10664-020-09829-4},
  copyright = {https://www.springer.com/tdm},
  langid = {english}
}

@article{sharafiSystematicLiteratureReview2015,
  title = {A Systematic Literature Review on the Usage of Eye-Tracking in Software Engineering},
  author = {Sharafi, Zohreh and Soh, Z{\'e}phyrin and Gu{\'e}h{\'e}neuc, Yann-Ga{\"e}l},
  year = 2015,
  month = nov,
  journal = {Information and Software Technology},
  volume = {67},
  pages = {79--107},
  issn = {09505849},
  doi = {10.1016/j.infsof.2015.06.008},
  langid = {english}
}

@inproceedings{busjahnEyeMovementsCode2015b,
  title = {Eye {{Movements}} in {{Code Reading}}: {{Relaxing}} the {{Linear Order}}},
  shorttitle = {Eye {{Movements}} in {{Code Reading}}},
  booktitle = {2015 {{IEEE}} 23rd {{International Conference}} on {{Program Comprehension}}},
  author = {Busjahn, Teresa and Bednarik, Roman and Begel, Andrew and Crosby, Martha and Paterson, James H. and Schulte, Carsten and Sharif, Bonita and Tamm, Sascha},
  year = 2015,
  month = may,
  pages = {255--265},
  publisher = {IEEE},
  address = {Florence, Italy},
  doi = {10.1109/ICPC.2015.36},
  isbn = {978-1-4673-8159-8},
  langid = {english}
}

@inproceedings{chorowskiAttentionBasedModelsSpeech2015,
  title = {Attention-{{Based Models}} for {{Speech Recognition}}},
  booktitle = {Advances in {{Neural Information Processing Systems}}},
  author = {Chorowski, Jan K and Bahdanau, Dzmitry and Serdyuk, Dmitriy and Cho, Kyunghyun and Bengio, Yoshua},
  year = 2015,
  volume = {28},
  publisher = {Curran Associates, Inc.}
}

@inproceedings{cordonnierRelationshipSelfAttentionConvolutional2019,
  title = {On the {{Relationship}} between {{Self-Attention}} and {{Convolutional Layers}}},
  booktitle = {International {{Conference}} on {{Learning Representations}}},
  author = {Cordonnier, Jean-Baptiste and Loukas, Andreas and Jaggi, Martin},
  year = 2019,
  month = sep,
  langid = {english}
}

@misc{gravesGeneratingSequencesRecurrent2014,
  title = {Generating {{Sequences With Recurrent Neural Networks}}},
  author = {Graves, Alex},
  year = 2014,
  month = jun,
  number = {arXiv:1308.0850},
  publisher = {arXiv},
  doi = {10.48550/arXiv.1308.0850},
  langid = {american}
}

@misc{gregorDRAWRecurrentNeural2015,
  title = {{{DRAW}}: {{A Recurrent Neural Network For Image Generation}}},
  shorttitle = {{{DRAW}}},
  author = {Gregor, Karol and Danihelka, Ivo and Graves, Alex and Rezende, Danilo Jimenez and Wierstra, Daan},
  year = 2015,
  month = may,
  number = {arXiv:1502.04623},
  publisher = {arXiv},
  doi = {10.48550/arXiv.1502.04623}
}

@inproceedings{sharifEyeTrackingStudy2010a,
  title = {An {{Eye Tracking Study}} on {{camelCase}} and Under\_score {{Identifier Styles}}},
  booktitle = {2010 {{IEEE}} 18th {{International Conference}} on {{Program Comprehension}}},
  author = {Sharif, Bonita and Maletic, Jonathan I.},
  year = 2010,
  month = jun,
  pages = {196--205},
  issn = {1092-8138},
  doi = {10.1109/ICPC.2010.41}
}

@inproceedings{youHardCodedGaussianAttention2020,
  title = {Hard-{{Coded Gaussian Attention}} for {{Neural Machine Translation}}},
  booktitle = {Proceedings of the 58th {{Annual Meeting}} of the {{Association}} for {{Computational Linguistics}}},
  author = {You, Weiqiu and Sun, Simeng and Iyyer, Mohit},
  editor = {Jurafsky, Dan and Chai, Joyce and Schluter, Natalie and Tetreault, Joel},
  year = 2020,
  month = jul,
  pages = {7689--7700},
  publisher = {Association for Computational Linguistics},
  address = {Online},
  doi = {10.18653/v1/2020.acl-main.687}
}

@misc{PCGrad,
  title = {Gradient Surgery for Multi-Task Learning},
  author = {Yu, Tianhe and Kumar, Saurabh and Gupta, Abhishek and Levine, Sergey and Hausman, Karol and Finn, Chelsea},
  year = 2020,
  month = dec,
  number = {arXiv:2001.06782},
  publisher = {arXiv},
  doi = {10.48550/arXiv.2001.06782},
  langid = {american}
}

@inproceedings{banerjeeMETEORAutomaticMetric2005,
  title = {{{METEOR}}: {{An}} Automatic Metric for {{MT}} Evaluation with Improved Correlation with Human Judgments},
  shorttitle = {{{METEOR}}},
  booktitle = {Proceedings of the {{ACL Workshop}} on {{Intrinsic}} and {{Extrinsic Evaluation Measures}} for {{Machine Translation}} and/or {{Summarization}}},
  author = {Banerjee, Satanjeev and Lavie, Alon},
  editor = {Goldstein, Jade and Lavie, Alon and Lin, Chin-Yew and Voss, Clare},
  year = 2005,
  month = jun,
  pages = {65--72},
  publisher = {Association for Computational Linguistics},
  address = {Ann Arbor, Michigan}
}

@inproceedings{linROUGEPackageAutomatic2004,
  title = {{{ROUGE}}: {{A}} Package for Automatic Evaluation of Summaries},
  shorttitle = {{{ROUGE}}},
  booktitle = {Text {{Summarization Branches Out}}},
  author = {Lin, Chin-Yew},
  year = 2004,
  month = jul,
  pages = {74--81},
  publisher = {Association for Computational Linguistics},
  address = {Barcelona, Spain}
}

@inproceedings{papineniBleuMethodAutomatic2002a,
  title = {Bleu: {{A}} Method for Automatic Evaluation of Machine Translation},
  shorttitle = {Bleu},
  booktitle = {Proceedings of the 40th {{Annual Meeting}} of the {{Association}} for {{Computational Linguistics}}},
  author = {Papineni, Kishore and Roukos, Salim and Ward, Todd and Zhu, Wei-Jing},
  editor = {Isabelle, Pierre and Charniak, Eugene and Lin, Dekang},
  year = 2002,
  month = jul,
  pages = {311--318},
  publisher = {Association for Computational Linguistics},
  address = {Philadelphia, Pennsylvania, USA},
  doi = {10.3115/1073083.1073135}
}

@misc{zhangBERTScoreEvaluatingText2020,
  title = {{{BERTScore}}: {{Evaluating}} Text Generation with {{BERT}}},
  shorttitle = {{{BERTScore}}},
  author = {Zhang, Tianyi and Kishore, Varsha and Wu, Felix and Weinberger, Kilian Q. and Artzi, Yoav},
  year = 2020,
  month = feb,
  number = {arXiv:1904.09675},
  publisher = {arXiv},
  doi = {10.48550/arXiv.1904.09675}
}

@misc{fartaleDisentanglingRecallReasoning2025,
  title = {Disentangling Recall and Reasoning in Transformer Models through Layer-Wise Attention and Activation Analysis},
  author = {Fartale, Harshwardhan and Kattamuri, Ashish and Raja, Rahul and Vats, Arpita and Prasad, Ishita and Moharir, Akshata Kishore},
  year = 2025,
  month = oct,
  number = {arXiv:2510.03366},
  publisher = {arXiv},
  doi = {10.48550/arXiv.2510.03366},
  langid = {american}
}

@misc{merulloTalkingHeadsUnderstanding2025a,
  title = {Talking Heads: {{Understanding}} Inter-Layer Communication in Transformer Language Models},
  shorttitle = {Talking {{Heads}}},
  author = {Merullo, Jack and Eickhoff, Carsten and Pavlick, Ellie},
  year = 2025,
  month = may,
  number = {arXiv:2406.09519},
  publisher = {arXiv},
  doi = {10.48550/arXiv.2406.09519}
}

@misc{skeanLayerLayerUncovering2025,
  title = {Layer by Layer: {{Uncovering}} Hidden Representations in Language Models},
  shorttitle = {Layer by {{Layer}}},
  author = {Skean, Oscar and Arefin, Md Rifat and Zhao, Dan and Patel, Niket and Naghiyev, Jalal and LeCun, Yann and {Shwartz-Ziv}, Ravid},
  year = 2025,
  month = jun,
  number = {arXiv:2502.02013},
  publisher = {arXiv},
  doi = {10.48550/arXiv.2502.02013},
  langid = {american}
}

@inproceedings{zhangInvestigatingLayerImportance2024,
  title = {Investigating Layer Importance in Large Language Models},
  booktitle = {Proceedings of the 7th {{BlackboxNLP Workshop}}: {{Analyzing}} and {{Interpreting Neural Networks}} for {{NLP}}},
  author = {Zhang, Yang and Dong, Yanfei and Kawaguchi, Kenji},
  year = 2024,
  pages = {469--479},
  publisher = {Association for Computational Linguistics},
  address = {Miami, Florida, US},
  doi = {10.18653/v1/2024.blackboxnlp-1.29},
  langid = {english}
}

@inproceedings{perry2023ccs,
  title = {Do Users Write More Insecure Code with {{AI}} Assistants?},
  booktitle = {Proceedings of the 2023 {{ACM SIGSAC Conference}} on {{Computer}} and {{Communications Security}}},
  author = {Perry, Neil and Srivastava, Megha and Kumar, Deepak and Boneh, Dan},
  year = 2023,
  month = nov,
  pages = {2785--2799},
  publisher = {ACM},
  address = {Copenhagen Denmark},
  doi = {10.1145/3576915.3623157},
  isbn = {979-8-4007-0050-7},
  langid = {english}
}

@misc{barke2023grounded,
  title = {Grounded Copilot: {{How}} Programmers Interact with Code-Generating Models},
  shorttitle = {Grounded {{Copilot}}},
  author = {Barke, Shraddha and James, Michael B. and Polikarpova, Nadia},
  year = 2022,
  month = oct,
  number = {arXiv:2206.15000},
  publisher = {arXiv},
  doi = {10.48550/arXiv.2206.15000}
}

@misc{liang2024icse,
  title = {A Large-Scale Survey on the Usability of {{AI}} Programming Assistants: {{Successes}} and Challenges},
  shorttitle = {A {{Large-Scale Survey}} on the {{Usability}} of {{AI Programming Assistants}}},
  author = {Liang, Jenny T. and Yang, Chenyang and Myers, Brad A.},
  year = 2023,
  month = sep,
  number = {arXiv:2303.17125},
  publisher = {arXiv},
  doi = {10.48550/arXiv.2303.17125}
}

@misc{fu2024tosem,
  title = {Security Weaknesses of Copilot-Generated Code in {{GitHub}} Projects: {{An}} Empirical Study},
  shorttitle = {Security {{Weaknesses}} of {{Copilot-Generated Code}} in {{GitHub Projects}}},
  author = {Fu, Yujia and Liang, Peng and Tahir, Amjed and Li, Zengyang and Shahin, Mojtaba and Yu, Jiaxin and Chen, Jinfu},
  year = 2025,
  month = feb,
  number = {arXiv:2310.02059},
  publisher = {arXiv},
  doi = {10.48550/arXiv.2310.02059}
}

@inproceedings{sabouri2025icse_trust,
  title = {Trust Dynamics in {{AI-assisted}} Development: {{Definitions}}, Factors, and Implications},
  shorttitle = {Trust {{Dynamics}} in {{AI-Assisted Development}}},
  booktitle = {2025 {{IEEE}}/{{ACM}} 47th {{International Conference}} on {{Software Engineering}} ({{ICSE}})},
  author = {Sabouri, Sadra and Eibl, Philipp and Zhou, Xinyi and Ziyadi, Morteza and Medvidovic, Nenad and Lindemann, Lars and Chattopadhyay, Souti},
  year = 2025,
  month = apr,
  pages = {1678--1690},
  publisher = {IEEE},
  address = {Ottawa, ON, Canada},
  doi = {10.1109/ICSE55347.2025.00199},
  isbn = {979-8-3315-0569-1}
}

@misc{choudhuri2025icse,
  title = {What Guides Our Choices? {{Modeling}} Developers' Trust and Behavioral Intentions towards {{GenAI}}},
  shorttitle = {What {{Guides Our Choices}}?},
  author = {Choudhuri, Rudrajit and Trinkenreich, Bianca and Pandita, Rahul and Kalliamvakou, Eirini and Steinmacher, Igor and Gerosa, Marco and Sanchez, Christopher and Sarma, Anita},
  year = 2024,
  month = dec,
  number = {arXiv:2409.04099},
  publisher = {arXiv},
  doi = {10.48550/arXiv.2409.04099}
}

@misc{lyu2025ase,
  title = {"my Productivity Is Boosted, but ..." Demystifying Users' Perception on {{AI}} Coding Assistants},
  author = {Lyu, Yunbo and Yang, Zhou and Shi, Jieke and Chang, Jianming and Liu, Yue and Lo, David},
  year = 2025,
  month = aug,
  number = {arXiv:2508.12285},
  publisher = {arXiv},
  doi = {10.48550/arXiv.2508.12285}
}

@article{alakmehPredictingCodeComprehension2024a,
  title = {Predicting {{Code Comprehension}}: {{A Novel Approach}} to {{Align Human Gaze}} with {{Code}} Using {{Deep Neural Networks}}},
  shorttitle = {Predicting {{Code Comprehension}}},
  author = {Alakmeh, Tarek and Reich, David and J{\"a}ger, Lena and Fritz, Thomas},
  year = 2024,
  month = jul,
  journal = {Proceedings of the ACM on Software Engineering},
  volume = {1},
  number = {FSE},
  pages = {1982--2004},
  issn = {2994-970X},
  doi = {10.1145/3660795},
  copyright = {https://creativecommons.org/licenses/by/4.0/},
  langid = {english}
}

@article{bansalModelingHumanAttention2023b,
  title = {Towards {{Modeling Human Attention}} from {{Eye Movements}} for {{Neural Source Code Summarization}}},
  author = {Bansal, Aakash and Sharif, Bonita and McMillan, Collin},
  year = 2023,
  month = may,
  journal = {Proceedings of the ACM on Human-Computer Interaction},
  volume = {7},
  number = {ETRA},
  pages = {1--19},
  issn = {2573-0142},
  doi = {10.1145/3591136},
  langid = {american}
}

@inproceedings{paltenghiThinkingDeveloperComparing2021,
  title = {Thinking like a Developer? {{Comparing}} the Attention of Humans with Neural Models of Code},
  shorttitle = {Thinking {{Like}} a {{Developer}}?},
  booktitle = {2021 36th {{IEEE}}/{{ACM International Conference}} on {{Automated Software Engineering}} ({{ASE}})},
  author = {Paltenghi, Matteo and Pradel, Michael},
  year = 2021,
  month = nov,
  pages = {867--879},
  publisher = {IEEE},
  address = {Melbourne, Australia},
  doi = {10.1109/ase51524.2021.9678712},
  copyright = {https://ieeexplore.ieee.org/Xplorehelp/downloads/license-information/IEEE.html},
  langid = {english}
}

@inproceedings{rodegheroImprovingAutomatedSource2014b,
  title = {Improving Automated Source Code Summarization via an Eye-Tracking Study of Programmers},
  booktitle = {Proceedings of the 36th {{International Conference}} on {{Software Engineering}}},
  author = {Rodeghero, Paige and McMillan, Collin and McBurney, Paul W. and Bosch, Nigel and D'Mello, Sidney},
  year = 2014,
  month = may,
  pages = {390--401},
  publisher = {ACM},
  address = {Hyderabad India},
  doi = {10.1145/2568225.2568247},
  isbn = {978-1-4503-2756-5},
  langid = {english}
}

@misc{zhangEyeMulatorImprovingCode2025,
  title = {{{EyeMulator}}: {{Improving Code Language Models}} by {{Mimicking Human Visual Attention}}},
  shorttitle = {{{EyeMulator}}},
  author = {Zhang, Yifan and Huang, Chen and Zhang, Yueke and Zhang, Jiahao and Li, Toby Jia-Jun and McMillan, Collin and Leach, Kevin and Huang, Yu},
  year = 2025,
  month = aug,
  number = {arXiv:2508.16771},
  publisher = {arXiv},
  doi = {10.48550/arXiv.2508.16771},
  langid = {american}
}

@inproceedings{zhang2020retrieval,
author = {Zhang, Jian and Wang, Xu and Zhang, Hongyu and Sun, Hailong and Liu, Xudong},
title = {Retrieval-based neural source code summarization},
year = {2020},
isbn = {9781450371216},
publisher = {Association for Computing Machinery},
address = {New York, NY, USA},
url = {https://doi.org/10.1145/3377811.3380383},
doi = {10.1145/3377811.3380383},
booktitle = {Proceedings of the ACM/IEEE 42nd International Conference on Software Engineering},
pages = {1385–1397},
numpages = {13},
location = {Seoul, South Korea},
series = {ICSE '20}
}

@inproceedings{alon2019code2seq,
      title={code2seq: Generating Sequences from Structured Representations of Code}, 
      author={Uri Alon and Shaked Brody and Omer Levy and Eran Yahav},
      year={2019},
      eprint={1808.01400},
      archivePrefix={arXiv},
      primaryClass={cs.LG},
      url={https://arxiv.org/abs/1808.01400}, 
}

@article{lu2021codexglue,
      title={CodeXGLUE: A Machine Learning Benchmark Dataset for Code Understanding and Generation}, 
      author={Shuai Lu and Daya Guo and Shuo Ren and Junjie Huang and Alexey Svyatkovskiy and Ambrosio Blanco and Colin Clement and Dawn Drain and Daxin Jiang and Duyu Tang and Ge Li and Lidong Zhou and Linjun Shou and Long Zhou and Michele Tufano and Ming Gong and Ming Zhou and Nan Duan and Neel Sundaresan and Shao Kun Deng and Shengyu Fu and Shujie Liu},
      year={2021},
      eprint={2102.04664},
      archivePrefix={arXiv},
      primaryClass={cs.SE},
      url={https://arxiv.org/abs/2102.04664}, 
}

@inproceedings{shi2022evaluation,
   title={On the evaluation of neural code summarization},
   url={http://dx.doi.org/10.1145/3510003.3510060},
   DOI={10.1145/3510003.3510060},
   booktitle={Proceedings of the 44th International Conference on Software Engineering},
   publisher={ACM},
   author={Shi, Ensheng and Wang, Yanlin and Du, Lun and Chen, Junjie and Han, Shi and Zhang, Hongyu and Zhang, Dongmei and Sun, Hongbin},
   year={2022},
   month=may, pages={1597–1608},
   collection={ICSE ’22} }

@inproceedings{gao2023makes,
   title={What Makes Good In-Context Demonstrations for Code Intelligence Tasks with LLMs?},
   url={http://dx.doi.org/10.1109/ASE56229.2023.00109},
   DOI={10.1109/ase56229.2023.00109},
   booktitle={2023 38th IEEE/ACM International Conference on Automated Software Engineering (ASE)},
   publisher={IEEE},
   author={Gao, Shuzheng and Wen, Xin-Cheng and Gao, Cuiyun and Wang, Wenxuan and Zhang, Hongyu and Lyu, Michael R.},
   year={2023},
   month=sep, pages={761–773} }

@article{fang2025enhanced,
author = {Fang, Minying and Yuan, Xing and Li, Yuying and Li, Haojie and Fang, Chunrong and Du, Junwei},
title = {Enhanced Prompting Framework for Code Summarization with Large Language Models},
year = {2025},
issue_date = {July 2025},
publisher = {Association for Computing Machinery},
address = {New York, NY, USA},
volume = {2},
number = {ISSTA},
url = {https://doi.org/10.1145/3728949},
doi = {10.1145/3728949},
journal = {Proc. ACM Softw. Eng.},
month = jun,
articleno = {ISSTA072},
numpages = {24}
}

@misc{sun2024source,
      title={Source Code Summarization in the Era of Large Language Models}, 
      author={Weisong Sun and Yun Miao and Yuekang Li and Hongyu Zhang and Chunrong Fang and Yi Liu and Gelei Deng and Yang Liu and Zhenyu Chen},
      year={2025},
      eprint={2407.07959},
      archivePrefix={arXiv},
      primaryClass={cs.SE},
      url={https://arxiv.org/abs/2407.07959}, 
}

@article{su2024distilled,
author = {Su, Chia-Yi and McMillan, Collin},
title = {Distilled GPT for source code summarization},
year = {2024},
issue_date = {May 2024},
publisher = {Kluwer Academic Publishers},
address = {USA},
volume = {31},
number = {1},
issn = {0928-8910},
url = {https://doi.org/10.1007/s10515-024-00421-4},
doi = {10.1007/s10515-024-00421-4},
journal = {Automated Software Engg.},
month = mar,
numpages = {26}
}

@inproceedings{mastropaolo2024evaluating,
author = {Mastropaolo, Antonio and Ciniselli, Matteo and Di Penta, Massimiliano and Bavota, Gabriele},
title = {Evaluating Code Summarization Techniques: A New Metric and an Empirical Characterization},
year = {2024},
isbn = {9798400702174},
publisher = {Association for Computing Machinery},
address = {New York, NY, USA},
url = {https://doi.org/10.1145/3597503.3639174},
doi = {10.1145/3597503.3639174},
booktitle = {Proceedings of the IEEE/ACM 46th International Conference on Software Engineering},
articleno = {218},
numpages = {13},
location = {Lisbon, Portugal},
series = {ICSE '24}
}

@article{virk2025calibration,
author = {Virk, Yuvraj and Devanbu, Premkumar and Ahmed, Toufique},
title = {Calibration of Large Language Models on Code Summarization},
year = {2025},
issue_date = {July 2025},
publisher = {Association for Computing Machinery},
address = {New York, NY, USA},
volume = {2},
number = {FSE},
url = {https://doi.org/10.1145/3729400},
doi = {10.1145/3729400},
journal = {Proc. ACM Softw. Eng.},
month = jun,
articleno = {FSE130},
numpages = {21}
}

@inproceedings{li2024machines,
author = {Li, Jiliang and Zhang, Yifan and Karas, Zachary and McMillan, Collin and Leach, Kevin and Huang, Yu},
title = {Do Machines and Humans Focus on Similar Code? Exploring Explainability of Large Language Models in Code Summarization},
year = {2024},
isbn = {9798400705861},
publisher = {Association for Computing Machinery},
address = {New York, NY, USA},
url = {https://doi.org/10.1145/3643916.3644434},
doi = {10.1145/3643916.3644434},
booktitle = {Proceedings of the 32nd IEEE/ACM International Conference on Program Comprehension},
pages = {47–51},
numpages = {5},
location = {Lisbon, Portugal},
series = {ICPC '24}
}

@inproceedings{mondal2023robust,
    title = "Robust Code Summarization",
    author = "Mondal, Debanjan  and
      Lodha, Abhilasha  and
      Sahoo, Ankita  and
      Kumari, Beena",
    editor = "Hupkes, Dieuwke  and
      Dankers, Verna  and
      Batsuren, Khuyagbaatar  and
      Sinha, Koustuv  and
      Kazemnejad, Amirhossein  and
      Christodoulopoulos, Christos  and
      Cotterell, Ryan  and
      Bruni, Elia",
    booktitle = "Proceedings of the 1st GenBench Workshop on (Benchmarking) Generalisation in NLP",
    month = dec,
    year = "2023",
    address = "Singapore",
    publisher = "Association for Computational Linguistics",
    url = "https://aclanthology.org/2023.genbench-1.5/",
    doi = "10.18653/v1/2023.genbench-1.5",
    pages = "65--75",
}

\end{document}